\documentclass[twocolumn,astrosymb]{aastex631}

\usepackage{amsmath}
\usepackage{natbib}
\usepackage{CJK}

\shorttitle{Faint Emission Lines in $z\sim2-3$ Galaxies}
\shortauthors{Strom et~al.}
\graphicspath{{./}{figures/}}

\begin{document}
\begin{CJK*}{UTF8}{gbsn}

\title{CECILIA: The Faint Emission Line Spectrum of $z\sim2-3$ Star-forming Galaxies}

\author[0000-0001-6369-1636]{Allison L. Strom}
\affiliation{Department of Physics and Astronomy, Northwestern University, 2145 Sheridan Road, Evanston, IL 60208, USA}
\affiliation{Center for Interdisciplinary Exploration and Research in Astrophysics (CIERA), Northwestern University, 1800 Sherman Avenue, Evanston, IL 60201, USA}

\author[0000-0002-8459-5413]{Gwen C. Rudie}
\affiliation{Carnegie Observatories, 813 Santa Barbara Street, Pasadena, CA 91101, USA}

\author[0000-0002-6967-7322]{Ryan F. Trainor}
\affiliation{Department of Physics and Astronomy, Franklin \& Marshall College, 637 College Avenue, Lancaster, PA 17603, USA}

\author[0000-0003-2680-005X]{Gabriel B. Brammer}
\affiliation{Cosmic Dawn Center (DAWN), Niels Bohr Institute, University of Copenhagen, Jagtvej 128, København N, DK-2200, Denmark}

\author[0000-0003-0695-4414]{Michael V. Maseda}
\affiliation{Department of Astronomy, University of Wisconsin, 475 N. Charter Street, Madison, WI 53706, USA}

\author[0009-0008-2226-5241]{Menelaos Raptis}
\affiliation{Department of Physics and Astronomy, Franklin \& Marshall College, 637 College Avenue, Lancaster, PA 17603, USA}

\author[0000-0002-0361-8223]{Noah S. J. Rogers}
\affiliation{Department of Physics and Astronomy, Northwestern University, 2145 Sheridan Road, Evanston, IL 60208, USA}
\affiliation{Center for Interdisciplinary Exploration and Research in Astrophysics (CIERA), Northwestern University, 1800 Sherman Avenue, Evanston, IL 60201, USA}
\affiliation{Minnesota Institute for Astrophysics, University of Minnesota, 116 Church St. SE, Minneapolis, MN, 55455}

\author[0000-0002-4834-7260]{Charles C. Steidel}
\affiliation{Cahill Center for Astronomy and Astrophysics, California Institute of Technology, MS 249-17, Pasadena, CA 91125, USA}

\author[0000-0003-4520-5395]{Yuguang Chen (陈昱光)}
\affiliation{Cahill Center for Astronomy and Astrophysics, California Institute of Technology, 1200 E California Blvd, MC249-17, Pasadena, CA 91125, USA}

\author[0000-0002-9402-186X]{David R. Law}
\affiliation{Space Telescope Science Institute, 3700 San Martin Drive, Baltimore, MD 21218, USA}

\correspondingauthor{Allison Strom}
\email{allison.strom@northwestern.edu}

\begin{abstract}
We present the first results from CECILIA, a Cycle 1 JWST NIRSpec/MSA program that uses ultra-deep $\sim30$~hour G235M/F170LP observations to target multiple electron temperature-sensitive auroral lines in the spectra of 33 galaxies at $z\sim1-3$. Using a subset of 23 galaxies, we construct two $\sim600$ object-hour composite spectra, both with and without the stellar continuum, and use these to investigate the characteristic rest-optical ($\lambda_{\rm rest}\approx5700-8500$~\AA) spectrum of star-forming galaxies at the peak epoch of cosmic star formation. Emission lines of eight different elements (H, He, N, O, Si, S, Ar, and Ni) are detected, with most of these features observed to be $\lesssim3\%$ the strength of H$\alpha$. We report the characteristic strength of three auroral features ([\ion{N}{2}]$\lambda5756$, [\ion{S}{3}]$\lambda6313$, and [\ion{O}{2}]$\lambda\lambda7322,7332$), as well as other semi-strong and faint emission lines, including forbidden [\ion{Ni}{2}]$\lambda\lambda7380,7414$ and permitted \ion{O}{1}~$\lambda8449$, some of which have never before been observed outside of the local universe. Using these measurements, we find $T_e$[\ion{N}{2}]~$=13630\pm2540$~K, representing the first measurement of electron temperature using [\ion{N}{2}] in the high-redshift universe. We also see evidence for broad line emission with a FWHM of $536^{+45}_{-167}$~km~s$^{-1}$; the broad component of H$\alpha$ is $6.01-28.31$\% the strength of the narrow component and likely arises from star-formation driven outflows. Finally, we briefly comment on the feasibility of obtaining large samples of faint emission lines using JWST in the future.
\end{abstract}

\section{Introduction}
\label{sec:intro}
The nebular emission lines originating in galaxies' star-forming regions are among the most powerful tools available for investigating the physical conditions in galaxies at all redshifts (see \citealt{kewley2019} for a review). Recombination lines of hydrogen provide constraints on star formation rates (SFRs) and reddening due to dust; weaker recombination lines of heavier elements such as oxygen are direct tracers of gas-phase enrichment; and collisionally-excited forbidden transitions of elements like oxygen, nitrogen, carbon, and sulfur are variously sensitive to electron temperature ($T_e$) and electron density ($n_e$), as well as the ionization state and enrichment of the photoionized gas.

Some of these lines, including H$\alpha$ at 6564~\AA\, and the [\ion{O}{3}]$\lambda\lambda4960,5008$ doublet, are bright enough to be detected out to large cosmological distances, even with relatively short exposure times. Indeed, over the last decade, ground-based near-infrared (NIR) spectrographs such as Magellan/FIRE \citep{simcoe2008,simcoe2010}, Keck/MOSFIRE \citep{mclean2010,mclean2012}, and VLT/KMOS \citep{sharples2013} have led to samples of 1000s of $z\sim2-3$ galaxies with measurements of such emission lines \citep[e.g.,][]{masters2014,steidel2014,shapley2015,sanders2015,wisnioski2015,strom2017}, allowing their metallicity, ionization and excitation properties, and gas density to be studied in comparable detail to large samples of $z\sim0$ galaxies \citep[e.g.,][]{kauffmann2003,brinchmann2004,tremonti2004,belfiore2015,mingozzi2020}. In the last year, JWST/NIRSpec and JWST/NIRCam grism observations have extended these efforts to even higher redshifts ($z\gtrsim3-6$) by enabling IR spectroscopy out to longer wavelengths \citep[e.g.,][]{kashino2023,kocevski2023,oesch2023,shapley2023,sun2023}.

Other lines---including metal recombination lines and the $T_e$-sensitive auroral lines of heavy elements, which are both key probes of chemical enrichment---are faint enough that they are not routinely detected even in spectra of nearby galaxies. Despite significant investment of observing time on some of the largest ground-based telescopes in the world, measurements of auroral [\ion{O}{3}]$\lambda4364$ were only possible for a handful of individual galaxies at $z\gtrsim2$ prior to the launch of JWST \citep{christensen2012,james2014,sanders2020}. Spectroscopic observations with JWST promise to yield unprecedented numbers of auroral emission line measurements in high-$z$ galaxies. The first analyses of the early release observations \citep[ERO;][]{pontoppidan2022} in the SMACS J0723.3–7327 field reinforced this expectation, with significant detections of [\ion{O}{3}]$\lambda4364$ in several $z\sim8$ galaxies \citep{arellano-cordova2022,schaerer2022,taylor2022,brinchmann2023,curti2023,katz2023,rhoads2023,trump2023,trussler2023}. At the same time, many of these studies reported conflicting gas-phase oxygen abundance (O/H) measurements in the same objects, and it was unclear how representative this early, very high-$z$ sample might be. Subsequent work has revisited the issue of auroral line detections in JWST observations of tens of high-$z$ galaxies (albeit primarily at low to moderate O/H), confirming suspicions that locally-calibrated metallicity diagnostics are likely unsuitable for the majority of high-$z$ galaxies \citep{laseter2023,sanders2023te}. To date, however, consensus regarding how best to leverage these measurements to, e.g., understand the overall distribution of chemical enrichment in the early universe has not yet been achieved.

In spite of these challenges, the community has collectively recognized the goal of using auroral line measurements and the resulting direct-method metallicities to construct more accurate methods of measuring high-$z$ galaxy enrichment \emph{in situ}. This is evidenced by the selection of three separate Cycle 1 JWST programs (PIDs 1879, 1914, and 2593) by the time allocation committee, with a total investment of over 150~hrs, or $\sim2.5$\% of all the GO time available in Cycle 1. Here, we report the first results from PID 2593, also known as CECILIA (Chemical Evolution Constrained Using Ionized Lines in Interstellar Aurorae; \citealt{strom2021}, data accessible via doi: \dataset[10.17909/x66z-p144]{https://doi.org/10.17909/x66z-p144}).

CECILIA was designed to measure auroral [\ion{S}{3}]$\lambda6313$ and [\ion{O}{2}]$\lambda\lambda7322,7332$ in the spectra of a carefully selected sample of $z\sim2-3$ star-forming galaxies, using $\sim30$~hr G235M/F170LP observations. Owing to the unique depth of these data, CECILIA is also able to detect myriad other lines in the galaxies' rest-optical spectra, some of which are stronger than any auroral emission line and, thus, more likely to be observed in more typical integration times with JWST. Consequently, it is important to understand the expected strength of these faint and semi-strong emission lines, in order to guide future studies using JWST, as well as with other current and future facilities.

The remainder of this letter focuses on two $\sim600$ object-hour rest-optical composite spectra of $z\sim2-3$ galaxies observed as part of the CECILIA survey, with the aim of providing an ``atlas" of the characteristic faint emission line spectrum of high-$z$ galaxies. We describe the CECILIA survey---including the galaxy sample, the JWST program, and the data reduction---in Section~\ref{sec:cecilia}. Section~\ref{sec:composite} outlines the construction of the composite spectra and their key features, with a more in-depth discussion of individual emission lines in Section~\ref{sec:emlines}. In Section~\ref{sec:conclusions}, we close with a summary of our findings and a brief discussion of implications for future observations of faint emission lines in $z\gtrsim2$ galaxies. Throughout the text, we refer to specific spectral features using their vacuum wavelengths.

\section{The CECILIA Survey}
\label{sec:cecilia}

\label{sec:sample}
\begin{figure*}
\centering
\includegraphics[width=0.67\textwidth]{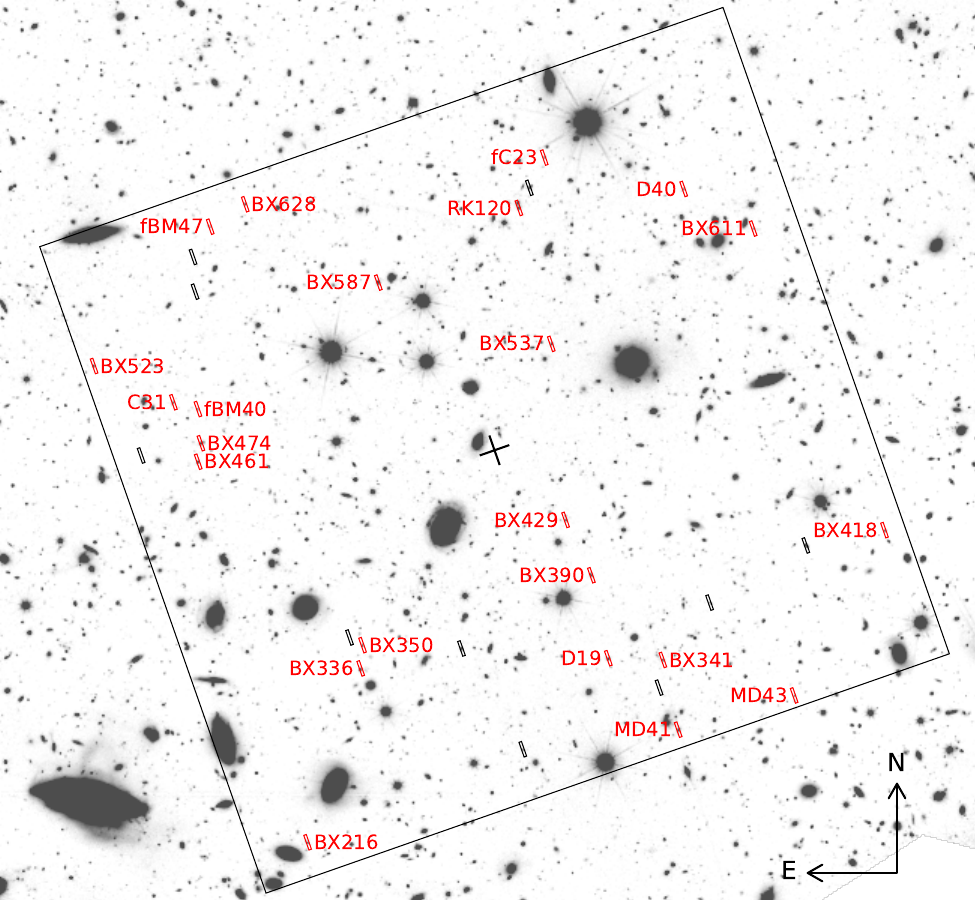}
\caption{The HST/WFC3 F140W image of Q2343+125 field targeted by CECILIA is shown in greyscale, with the approximate 3.6$'$$\times$3.4$'$ NIRSpec field of view overlaid in black. The observed slits are magnified by $\sim$3$\times$ for visibility, with red slits and names indicating the 23 sources included in this work.}
\label{fig:mosaic}
\end{figure*}

The principal goal of CECILIA is to measure multiple faint rest-optical auroral lines in the spectra of $z\sim2-3$ galaxies, which can then be used to calibrate new high-$z$ metallicity diagnostics. Some of the galaxies observed as part of CECILIA have preexisting rest-optical spectra obtained using Keck/MOSFIRE \citep{steidel2014,trainor2015,strom2017}, but even the strongest auroral lines are not routinely detected for individual galaxies in deep ($\sim8-10$~hr) observations. Although JWST/NIRSpec provides greater sensitivity and spectral coverage than ground-based NIR spectrographs, achieving this goal still pushes the limits of the observatory. To make the best use of JWST observing time, we first used detailed photoionization models and existing ground-based rest-ultraviolet (UV) and rest-optical spectra of the same galaxies to robustly predict the auroral line strengths. We then used these predictions together with the (pre-flight) JWST Exposure Time Calculator (ETC)\footnote{https://jwst.etc.stsci.edu/} to identify the depth needed to detect the auroral lines in individual galaxies. Below, we describe the parent galaxy sample, the emission line predictions, the design of the NIRSpec program, including exposure time requirements and microshutter assembly (MSA) design, and the reduction of the JWST data.

\subsection{Parent Galaxy Sample and Field Selection}

CECILIA targets galaxies drawn from the Keck Baryonic Structure Survey \citep[KBSS;][]{steidel2010,rudie2012,trainor2015,strom2017}. KBSS is a large spectroscopic survey of UV-color and narrow-band Ly$\alpha$ selected galaxies in 15 fields. The survey includes deep $J$, $H$, and $K$ NIR (rest-optical) spectroscopy from Keck/MOSFIRE, deep optical (rest-UV) spectroscopy from Keck/LRIS, and imaging in $U_n$ through $K_s$, \textit{HST}/WFC3 F140W, F160W, \textit{Spitzer} IRAC Ch$1-4$ and MIPS 24 $\mu$m, and narrow-band (rest-frame) Ly$\alpha$ filters. In total, KBSS comprises $\sim3500$ $z\sim1-3$ galaxies with confirmed (rest-UV or rest-optical) spectroscopic redshifts and rich multiwavelength data. Thanks to the dense sampling of galaxies with known redshifts and high-quality rest-optical spectroscopy in KBSS, targeting one of the survey fields allows us to efficiently prioritize those galaxies predicted to yield auroral line detections, where we can also ensure that all of the emission lines required to determine $T_e$ and direct-method metallicity are accessible.

The primary targets for CECILIA are UV-color-selected galaxies for which detecting the auroral [\ion{S}{3}]$\lambda6313$ line is feasible, as this is the faintest line required to achieve the primary goals of the program (see Section~\ref{sec:predictions}). We focused on galaxies with $2.10\leq z\leq2.68$, where all of the required emission lines fall within the G235M, G395M, and ground-based NIR spectral bandpasses. Galaxies were prioritized if they have: somewhat higher redshifts ($z>2.3$), where NIRSpec is more sensitive at the observed wavelength of [\ion{S}{3}]$\lambda6313$; smaller than average sizes, increasing the auroral surface brightness; high SED-based SFRs ($> 24~\rm M_\odot ~\rm yr^{-1}$); and/or large observed nebular [\ion{O}{3}]$\lambda5008$ or H$\alpha$ line fluxes ($> 7.0 \times 10^{-17}$ erg s$^{-1}$ cm$^{-2}$). All of these properties increase the ease of detection with the NIRSpec/MSA. The highest priority targets were galaxies with detailed emission line models (Section~\ref{sec:predictions}) whose predicted auroral line surface brightnesses exceeded the detection threshold of the planned observations; galaxies with models predicting non-detections were down weighted. Narrow-band selected Ly$\alpha$ emitters (LAEs) from \citet{trainor2016} with spectroscopic detections of Ly$\alpha$ and [\ion{O}{3}]$\lambda5008$ or H$\alpha$ were also prioritized as a way of extending the galaxy sample to lower stellar masses (M$_{\ast}$) and SFRs.

Of the 15 KBSS fields, we selected the Q2343+125 field due to its high density of high-priority sources and large catalog of LAEs at $z\approx2.55$ with spectroscopic redshifts. Further, this field also has an existing HST/WFC3 F140W mosaic (Figure~\ref{fig:mosaic}) that provided both the precision astrometry required for mask design and the galaxy size measurements needed for target prioritization---without requiring additional \mbox{(pre-)imaging} from space using JWST or HST.

The CECILIA JWST/NIRSpec observations contain a total sample of 34 galaxies.\footnote{One target, Q2343-D27, appears to be a $z=0.0890$ interloper, based on the JWST/NIRSpec observations.} We include 23 of these objects here (Figure~\ref{fig:mosaic}), omitting the Ly$\alpha$-selected galaxies that do not have secure SED models (4 galaxies), galaxies at $z<2$ (4 galaxies), and sources that were severely impacted by shutter failures in the NIRSpec/MSA (3 galaxies). The final sample is largely typical of KBSS galaxies, with $\langle z\rangle=2.4$, masses spanning log(M$_{\ast}$/M$_{\odot})=8.5-10.7$ and a median value of log(M$_{\ast}$/M$_{\odot})=9.7$ (assuming a \citealt{chabrier2003} stellar initial mass function). Based on H$\alpha$ and H$\beta$ measurements from ancillary MOSFIRE spectra, the included galaxies have SFRs ranging from $16-42$~M$_{\odot}$~yr$^{-1}$, with a median SFR$_{\rm H\alpha}=21$~M$_{\odot}$~yr$^{-1}$. These are slightly lower than median values reported in \citet{strom2017}, which were determined in the same manner, but similar in terms of M$_{\ast}$ to the subsample of KBSS galaxies used to construct the deep ``LM1" composite in \citet{steidel2016}.

\begin{figure}
\centering
\includegraphics[trim=0 0.55in 0 0,clip]{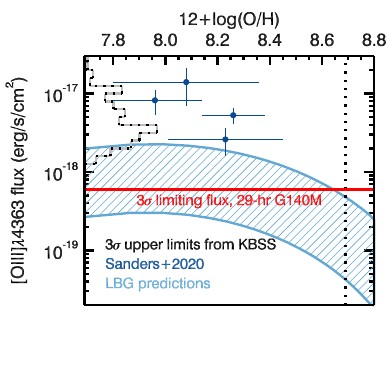}\\
\vspace{-0.04in}
\includegraphics[trim=0 0.13in 0 0.36in,clip]{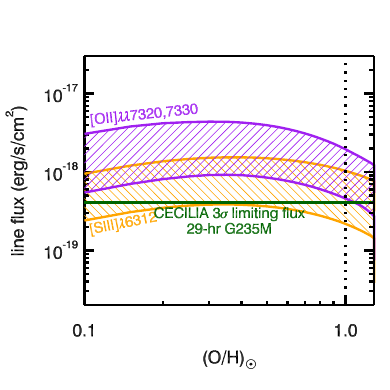}
\caption{The hatched regions show the model predictions for auroral lines in the rest-optical spectra of typical $z\sim2-3$ galaxies, separated on the basis of whether they fall in the G140M bandpass (top panel) or the G235M bandpass (bottom panel). The width of the hatched regions reflects the typical range of ionization parameter $U$ in high-$z$ galaxies. The predicted line fluxes for [\ion{O}{3}]$\lambda4364$ (blue hatched region) are $\sim1$~dex fainter than the depth of typical ground-based spectra of individual galaxies, represented by the distribution of $3\sigma$ upper limits on [\ion{O}{3}]$\lambda4364$ from KBSS (dashed black histogram). \citet{sanders2020} reported four ground-based detections of unlensed [\ion{O}{3}]$\lambda4364$ (blue points, shifted up by 0.24~dex to match the photoionization model abundance scale), but these galaxies appear atypical. Estimates using the pre-flight ETC indicated that detecting [\ion{O}{3}]$\lambda4364$ in a representative sample of high-$z$ galaxies would be prohibitively expensive; the $3\sigma$ limiting line flux of $6\times10^{-19}$~erg~s$^{-1}$~cm$^{-2}$ achievable in a combined 29~hr G140M exposure (red line) probes $\lesssim30\%$ of the $z\sim2-3$ sample from \citet{strom2018}. Fortunately, the typical predicted line fluxes for the sum of the [\ion{O}{2}]$\lambda\lambda7322,7332$ lines (purple hatched region) and [\ion{S}{3}]$\lambda6313$ (orange hatched region) could be detected for galaxies with a wider range of $U$ and O/H in the same exposure time, due to the higher sensitivity of NIRSpec in G235M. A $3\sigma$ limiting line flux of $4.1\times10^{-19}$~erg~s$^{-1}$~cm$^{-2}$ reaches $\sim90\%$ of typical galaxies.}
\label{fig:auroral}
\end{figure}

\subsection{Emission Line Predictions}
\label{sec:predictions}

The expected strengths of the auroral emission lines targeted by CECILIA were determined using photoionization models designed to reconcile the rest-UV and rest-optical spectra of $z\sim2-3$ galaxies. We used a combination of the Binary
Population and Spectral Synthesis models \citep[BPASSv2;][]{stanway2016,eldridge2017} and Cloudy photoionization models \citep[Cloudy13;][]{ferland2013} to predict line strengths as a function of gas-phase metallicity (O/H). We matched the model parameters (gas density $n_{\rm H}$, stellar Fe/H, and ionization parameter $U$) to the properties of $z\sim2-3$ KBSS galaxies reported by \citet{strom2018}, which are consistent with the values reported for other $z\sim2-3$ samples \citep[e.g.,][]{topping2020}. The model outputs were then converted to line fluxes using a representative range of SFRs and dust extinction.

Figure~\ref{fig:auroral} presents the predictions for three of the brightest rest-optical auroral emission lines as a function of O/H, with the width of the hatched regions corresponding to the typical range of $U$ in high-$z$ galaxies; lower ionization galaxies have fainter lines at fixed metallicity. Re-calibrating strong line metallicity diagnostics and photoionization models at $z\gtrsim2$ requires measuring auroral lines in galaxies spanning both O/H and $U$, as both directly influence the strength of nebular emission lines. The top panel in Figure~\ref{fig:auroral} shows the steep decline in [\ion{O}{3}]$\lambda4364$ with increasing O/H and implies a limited ability to detect typical $z\sim2$ galaxies at high O/H and/or low $U$ using JWST/NIRSpec, even with long exposures in G140M.

In contrast, the bottom panel of Figure~\ref{fig:auroral} shows that a $3\sigma$ line flux sensitivity of 4.1$\times10^{-19}$~erg~s$^{-1}$~cm$^{-2}$ in G235M (corresponding to a total $\approx30$~hr exposure time using the pre-flight ETC; see Section~\ref{sec:exposuretime}) enables the detection of [\ion{S}{3}]$\lambda6313$ \emph{and} [\ion{O}{2}]$\lambda\lambda7322,7332$ at virtually all $U$, even in galaxies with relatively high gas-phase O/H. It is comparatively easier to detect [\ion{S}{3}]$\lambda6313$ and [\ion{O}{2}]$\lambda\lambda7322,7332$ not only because they are predicted to be intrinsically brighter than [\ion{O}{3}]$\lambda4364$ in the same galaxies, but also because of the increasing sensitivity of JWST/NIRSpec at longer wavelengths. On the basis of these predictions, we elected to obtain deep spectra of galaxies in a \emph{single} configuration, in order to maximize the overall number of auroral lines detected for individual galaxies with a range of O/H and $U$.

\subsection{JWST/NIRSpec Program Design}
\label{sec:observations}

To optimize the efficiency of the JWST program, we generated a large grid of ETC simulations spanning a range of galaxy sizes, limiting line fluxes, MSA centering constraints, and redshifts, as well as a comparable grid of MSA Planning Tool (MPT) simulations that considered the full range of available centering constraints. In this section, we describe the most salient elements of the program design.

\subsubsection{Exposure time requirements}
\label{sec:exposuretime}
NIRSpec G235M observations of [\ion{S}{3}]$\lambda6313$ and [\ion{O}{2}]$\lambda\lambda7322,7332$ in CECILIA galaxies were modeled using an exponential surface brightness profile (Sersic index $n=1$) with a projected semi-major axis of 0\farcs26 and an axis ratio of $b/a=0.6$, consistent with the measured morphologies and median sizes of galaxies in our parent sample \citep{law12}.

Pre-flight ETC simulations showed that reaching the required $3\sigma$ limiting line flux of $4.1\times10^{-19}$~erg~s$^{-1}$~cm$^{-2}$ for a median-sized galaxy at $z=2.3$ at the edge of the midpoint tolerance (see Section~\ref{sec:msa}) required 29.5 hours of exposure time (20 groups $\times$ 6 integrations $\times$ 12 exposures) using NRS IRS2 readouts. Our exposure time calculations assumed ``MSA Full Shutter Extraction'' and assumed we would need pixel-level subtraction from A-B pairs. As we discuss below in Section~\ref{sec:bkgsub}, we have instead implemented a global background model drawn from slits across the full MSA, which reduces the overall noise in the final combined data compared to the conservative assumptions in our original calculations. For the majority of the sources in our catalog, ETC calculations demonstrated that some of the background region in each spectrum would be contaminated with light from the source, and the derived exposure time requirements took this effect into account.

\subsubsection{MSA design}
\label{sec:msa}

The MSA configuration is central to the success of CECILIA, and considerable experience with ground-based multi-object mask design led us to conduct extensive trials using different mask parameters in the MSA Planning Tool (MPT). We experimented with all possible centering constraints, dithering and nodding options, and three- and five-slitlet length slits. We ran trial masks spanning the full range of allowable PAs, using small steps in both position and PA to understand the sensitivity of the optimal configuration to changes in PA. Based on more than 100 runs of the MPT considering more than 70 million unique configurations, we determined that many PAs have $<$~60\% as many high-priority targets as the best masks.

We optimized the MSA centering constraint, which trades exposure time against sample size, by considering a grid of ETC and MPT runs. Our ETC calculations spanned the full redshift range of source galaxies and sizes ranging between the 1st and 3rd quartile of the KBSS size distribution. We considered the S/N penalty for galaxies at maximal offset in the dispersion direction for each of the three possible centering restrictions.\footnote{The assumed offsets in the ETC are 0\farcs063 for ``constrained,'' 0\farcs076 for ``midpoint,'' and 0\farcs099 for ``entire open shutter.'' These values use the specified margin and the pitch of the MSA shutters, thus accounting for the current uncertainty in the width of the MSA bars.} For galaxies with the median size in our sample, the S/N penalties compared to a perfectly centered target are $7-13$\% for ``constrained,'' $11-19$\% for ``midpoint,'' and $14-26$\% for ``entire open shutter,'' where the reported ranges represent different relative angles between the short axis of the slit and the major axis of the galaxies. MPT runs showed that ``constrained'' configurations allowed for only 60\% of the high-priority targets to be placed on a mask compared to the ``midpoint'' criteria. Relaxing the centering further via ``entire open shutter'' constraint only increased the number of high-priority targets by 7\%. Therefore, we selected the ``midpoint'' centering constraint for CECILIA observations.

We designed custom software that processed the MPT MSA configurations to check the wavelength coverage\footnote{The post-flight version of MPT now has the ability to output the wavelength coverage of individual slitlets.} (using MSAViz\footnote{\url{https://github.com/spacetelescope/msaviz}}) and confirmed that primary targets assigned to a slit on the MSA would have spectral coverage of the required auroral and nebular lines. This software also considers the known emission line properties, M$_{\ast}$, and SFRs of target galaxies, which we used to select a final mask configuration that appropriately sampled the parent sample to enable an effective metallicity calibration. We selected a default three-shutter slitlet shape with a three-point nod pattern within the slitlet.

Upon scheduling, we were assigned a Aperture Position Angle, APA~=~20.0, with values from $18.5<$~APA~$<20.0$ able to be accommodated within the scheduling window. At this point, we completed a second set of MPT simulations, including PA steps of 0.1 degree and $0\farcs025-0\farcs01$ position steps to optimize the PA and final mask. We did not reach convergence,\footnote{We define a converged mask as one where the same optimal mask is returned even when the step size is decreased.} even with angle and position steps much finer than suggested by JDox, suggesting that significant computational resources would be required to fully optimize NIRSpec/MSA observations. Based on our simulations, we ultimately selected an APA~=~19.3. Over the 1.5 degree range allowable within the plan window, the MPT resulted in more than a 30\% variation in the number of high-priority targets, and we advocate for conducting similar PA optimization to maximize the efficiency of other NIRSpec/MSA programs with low to moderate density of high priority targets. 

Following the selection of pointing and PA, we ran MPT with an expanded catalog to (1) check for contamination in any of the shutters known to be stuck open as of June 2022 and (2) open shutters on dark regions of the sky to sample the background light across the field. These sky slitlets are described in our modeling of the global sky background in Section~\ref{sec:bkgsub}. 
Once the automated MSA configuration was determined by MPT, the solution was hand-edited using the MSA configuration editor to (1) elongate slits for high priority targets where possible, (2) add more background shutters close to high priority targets to better sample relevant wavelength or field-position changes in the background, and (3) add high priority targets that did not meet our centering constraints but could be placed on a mask without conflicting with other high priority targets.
The final MSA design included 34 sources, 23 of which are included in the stacked spectra presented in this letter.

\subsection{JWST/NIRSpec Data Reduction}
\label{sec:reduction}

The uncalibrated raw G235M data (\texttt{uncal}) frames were processed using the \texttt{jwst\_level1} pipeline in the \texttt{grizli}\footnote{\url{https://github.com/gbrammer/grizli}} package version 1.8.9 from \citet{grizli}. The level 1 pipeline in \texttt{grizli} uses the \texttt{calwebb} Science Calibration Pipeline (\citealt{calwebb_v1.10.0}, version 1.10.0, \texttt{CRDS\_CONTEXT} $=$ jwst\_1100.pmap) for the \texttt{group\_scale} correction, initial flagging of bad pixels and saturated pixels, bias subtraction (including corrections to the bias using reference pixels), as well as corrections for detector linearity and persistence and subtraction of the dark current. Following these \texttt{calwebb} steps, clusters of pixels affected by snowball cosmic ray events were flagged and the ramp fit was calculated, including additional processing to detect and remove the effects of cosmic rays and detector defects. Finally, the gain correction was applied, resulting in the level 1 processed \texttt{rate} files. 

Next, the level 1 processed files were corrected for correlated read noise, which manifests as vertical banding in the \texttt{rate} files. This $1/f$ noise, driven by small temperature variations in the ASIC readout electronics, was modeled and removed using the \texttt{NSClean} algorithm from \citet{rauscher2023}. \texttt{NSClean} requires the user to create a mask that identifies areas on each of the two NIRSpec detectors that are unilluminated by source light; these areas are thus relatively clean tracers of the correlated readout noise. We tested many different mask design strategies in order to remove as much of the large-scale vertical banding as possible while also limiting the introduction of additional high-frequency noise, which we found to be a side effect of the \texttt{NSClean} algorithm in many cases. We determined the most effective masks for our program omitted entire rows of pixels in the rectified full-detector image if any portion of that row was illuminated. Mask designs that omitted only the limited range of pixels that were illuminated in a given row resulted in higher levels of high-frequency noise being introduced in the regions of the detectors that were illuminated by source light. 

Following the $1/f$ noise correction by \texttt{NSClean}, we applied the preprocessing routine steps from \texttt{msaexp}\footnote{\url{https://github.com/gbrammer/msaexp}} version 0.6.11 from \citet{msaexp}, aside from the $1/f$ noise correction. This routine repeated the search for snowballs and additional detector defects, which were also masked. We applied a bias offset correction calculated from the median of unilluminated pixels in each frame and rescaled the read noise array associated with each exposure so that it reflected the distribution of the same unilluminated pixels.

Next, we used \texttt{msaexp} to call the \texttt{calwebb\_spec2} JWST Science Calibration Pipeline (\citealt{calwebb_v1.10.0}, version 1.10.0), which computed the world coordinate system reference frame for the data (including the wavelength calibration), extracted the individual 2D spectra for each slit, and flat-fielded each 2D spectral cutout. Each spectral cutout was corrected for path loss assuming the sources uniformly illuminate the slit (i.e., using the \texttt{PATHLOSS\_UN} correction). Note that the current \texttt{calwebb\_spec2} pipeline does not apply path loss corrections for slits more than three shutters in length, of which there are many in CECILIA, for reasons described in Section~\ref{sec:msa}). Thus, we modified the pipeline to apply the uniform source path loss correction to all slits. The pipeline correction for the bar shadows produced by the discretized MSA slitlets was then applied to the data, although, as described in Section~\ref{sec:bkgsub}, the pipeline correction left residual bar shadows on the data and background illumination. The \texttt{calwebb\_spec2} \texttt{photom} step then provided a final correction to the photometric calibration of the data, resulting in flux-calibrated 2D spectra for each slit and exposure. Finally, we used the \texttt{msaexp} \texttt{drizzle} routine to resample the individual 2D spectra onto a common rectified pixel grid and combine the exposures for each slit with outlier rejection, using a threshold of 100. 

\subsubsection{Background subtraction and extraction}
\label{sec:bkgsub}

To correct the data for background light, we opted to use a full-MSA background solution, rather than a paired exposure differencing algorithm, for several reasons. First, subtracting a global background model maximizes the S/N in the final spectra by excluding the shot noise that would be added by using a low-S/N measure of the background from single adjoining shutters. Second, the CECILIA targets are extended objects with light from each galaxy contaminating the shutters above and below the primary shutter. As such, the typical background algorithms that directly subtract the detected signal above and below the primary shutter inevitably subtract some source light as well. This over-subtraction poses a particular issue at the wavelengths of bright emission lines, which frequently extend well beyond the typical 0\farcs6 dither spacing of our observations in the background-subtracted 2D spectra. Finally, as described below we found that a single global background model provided a good description of the background across the field, while also enabling useful checks on the systematics of our observations.

We constructed the global background model by combining data from all the illuminated shutters in the MSA. Each rectified and drizzled 2D science spectrum was masked to omit rows corresponding to continuum emission from target galaxies or from other sources identified in the slit. Pixels illuminated by extended emission lines were also masked. The full set of masked science spectra (including those from dedicated sky slitlets, which were not masked unless they included coincidental sources) were then median combined into a single 2D background model. The 2D background was averaged in the spectral direction in order to model the residual bar shadows that were not fully corrected by the \texttt{calwebb\_spec2} pipeline, and these residual bar shadows were then removed from the 2D background model. We then averaged the resulting 2D background model in the spatial direction, weighting each pixel by the number of spectra contributing to the 2D median at the corresponding point in order to construct a 1D average background model as a function of wavelength.

As a cross-check on the consistency of our global background model, we created similar models from subsets of the observed slits grouped by their position on the sky (quartiles in right ascension and declination), on the MSA (quadrants 1, 2, 3, and 4), as well as by separating the portions of spectra falling on each detector (NRS1 and NRS2). The estimated 1D background was consistent across the field, but we found a small additive offset\footnote{The source of this offset is not clear, but we speculate that it stems from inconsistencies in the bias estimated from each detector.} between the NRS1 and NRS2 detectors. We therefore applied a compensatory offset to the portions of each slit's recorded background spectrum falling on NRS2 before averaging the data from all slits to create the global background model.

Before subtracting the background from each 2D science spectrum, a slit-specific version of the global background mode was created that accounts for the NRS2 offset on the portions of that spectrum that fall on NRS2. This 1D model is then subtracted from each 2D science spectrum. 
 Notably, the background model we derived is generally consistent with the predictions of the JWST Background Tool\footnote{\url{https://jwst-docs.stsci.edu/jwst-other-tools/jwst-backgrounds-tool}} (JBT) for our observations. However, a small additive offset is required to make JBT prediction match the normalization of our empirical background model, and there are a number of small-scale spectra features in the empirical background that are unresolved or not included in the JBT spectrum.

Optimal extraction of the 1D spectra was performed using routines from \texttt{msaexp}. A spatial profile of the continuum emission for each background-subtracted 2D spectrum was created by averaging along the wavelength dimension after weighting by the pipeline-produced 2D weight mask and applying a sigma-clipping algorithm to mask contaminated pixels and bright emission lines. An analogous spatial profile of the nebular emission was also created for each source by averaging the 2D spectrum over small wavelength ranges centered at the locations of bright emission lines. Each resulting 1D spatial profile was then fit independently with a Gaussian model. The resulting fits were typically similar for the continuum and emission line profiles, with the median profile being 20\% wider for the emission lines than the continuum. We used the Gaussian emission-line spatial model to provide the weights for the optimal extraction, except in one case where the continuum profile was used owing to a visibly-poor fit to the emission lines.
\newline

Despite the efforts described above, there are still unresolved issues in the data reduction resulting from known issues with JWST data products,\footnote{\url{https://jwst-docs.stsci.edu/jwst-calibration-pipeline-caveats/known-issues-with-jwst-data-products}} including uncertain variations in the spectral response as a function of slit position. Likewise, the unexplained additive offset between the NRS1 and NRS2 detectors, the residual bar shadows in the pipeline-processed 2D spectra, and the disagreement between our estimated background and the JBT predictions suggest that there are systematic effects (perhaps related to detector bias) that are incorrectly handled by the current pipeline tools and have uncertain downstream effects. While these uncertainties are not tolerable for the primary goal of CECILIA---precise abundance determinations of individual galaxies---we expect the stacking and normalization procedures described in Section \ref{sec:composite} likely mitigate any systematic effects on our composite spectra.

\section{\texorpdfstring{The Characteristic Rest-optical Spectrum of $\langle z\rangle \sim 2.4$ Galaxies}{The Characteristic Rest-optical Spectrum of z~2.4 Galaxies}}
\label{sec:composite}

\begin{figure*}
\centering
\includegraphics[width=16.5cm]{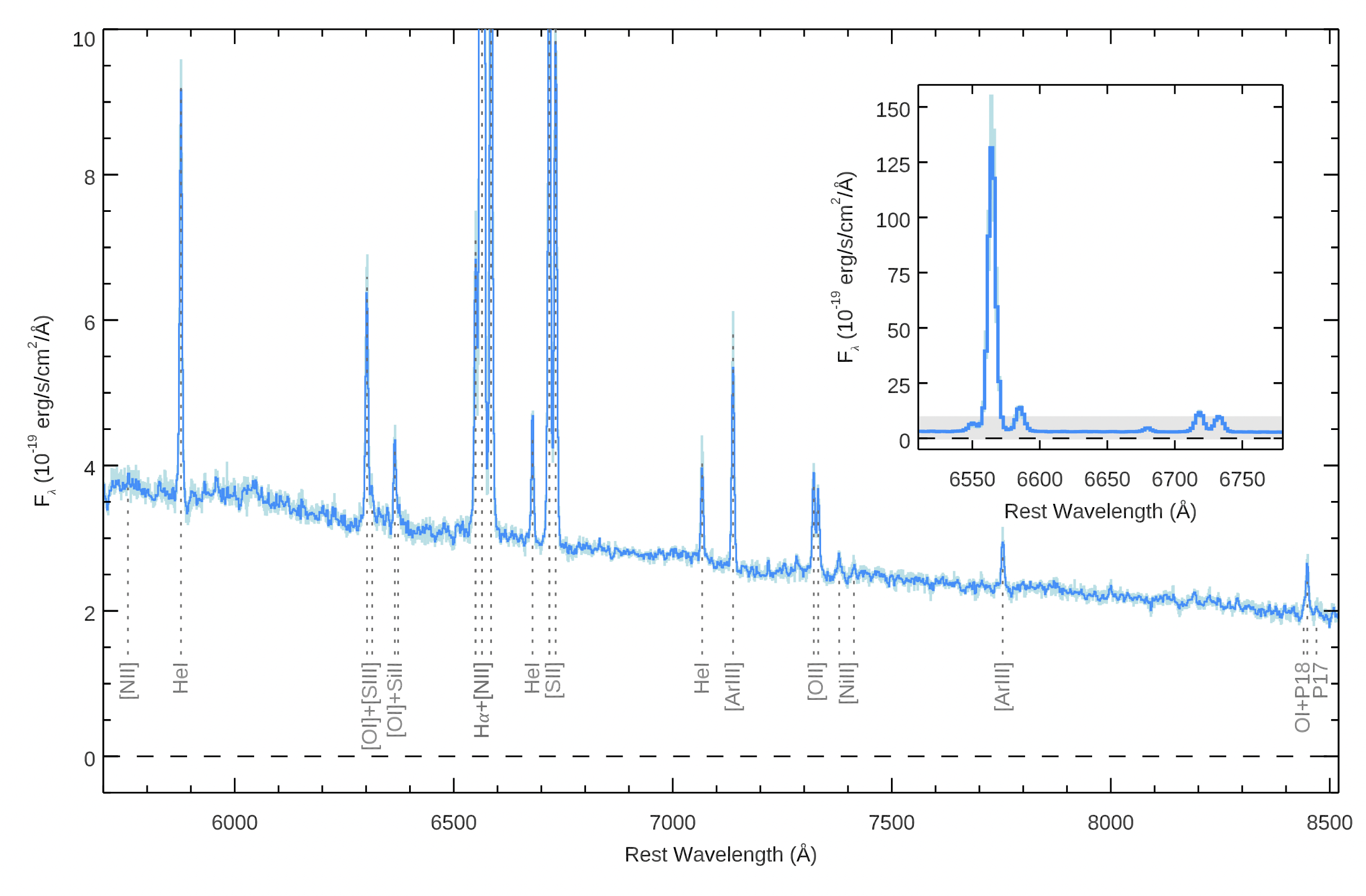}
\caption{The median-combined composite spectrum for the CECILIA sample is shown by the medium blue line, where the individual galaxy spectra are scaled by the observed continuum at $6800-7000$~\AA\, before being combined. The stack is then re-scaled so that the continuum in the same wavelength interval matches the median rest-frame continuum for the constituent galaxies. The 68\% CI for the composite is indicated by the light blue shading, and detected emission lines of eight different elements (H, He, N, O, Si, S, Ar, and Ni) are identified by dotted grey lines. The inset panel shows a zoomed-out version of the composite, centered on H$\alpha$, [\ion{N}{2}]$\lambda\lambda6550,6585$, and [\ion{S}{2}]$\lambda\lambda6718,6733$, which are the only lines routinely observed in ground-based observations and shallower JWST spectra of individual high-$z$ galaxies; the grey shaded region indicates the flux range shown in the full figure, where many fainter lines are visible.}
\label{fig:straight_stack}
\end{figure*}

CECILIA contains some of the deepest spectra obtained during Cycle 1, with $\sim30$~hr observations of individual galaxies using the NIRSpec/MSA and the G235M/F170LP configuration. These data offer a unique opportunity to investigate the spectra of high-$z$ star-forming galaxies, revealing features that have long remained out of reach of ground-based observations. Given the uncertainties in the data reduction at the present time, we use composite spectra as a tool to investigate the nebular emission lines observed in our data. We have two principal aims: (1) to illustrate the archetypal red rest-optical ($\lambda_{\rm rest}\approx5700-8500$~\AA) spectrum of a $z\sim2$ galaxy and (2) determine the typical range of emission line strengths. To achieve these goals, we construct two composite spectra, one including the stellar continuum and one only including the nebular emission. In this section, we describe how the two composite spectra are created, as well as their key features.

\subsection{The Total Composite Spectrum}

The flux scale of each reduced 1D spectrum is adjusted by comparing the observed continuum with the best-fit spectral energy distribution (SED) model of the same galaxy. This strategy has become common practice in analyzing JWST spectra of high-$z$ galaxies as a way of accounting for uncertainties in the flux calibration. Specifically, we mask regions of the spectra with large deviations from the median flux level ($\geq2\times$ the median absolute deviation), which excludes not only strong emission lines but also any serious artifacts remaining in the data due to bad pixels and cosmic rays. We then use a low-order polynomial to define a multiplicative ``slit loss" function for each object that forces the observed continuum to match the best-fit SED.

After this additional flux correction step, the spectra are shifted into the rest frame and normalized by the median observed continuum flux in the region between $\lambda_{\rm rest}=6800-7000$~\AA, where there are no emission lines; this portion of the spectrum is also approximately centered with respect to the auroral [\ion{S}{3}]$\lambda6313$ and [\ion{O}{2}$]\lambda\lambda7322,7332$ lines. The spectra are then interpolated onto a common rest-frame wavelength array and median-combined. The final stack is subsequently rescaled to match the median rest-frame continuum of all the constituent galaxies between $\lambda_{\rm rest}=6800-7000$~\AA. Uncertainties are estimated by generating 1000 bootstrap-resampled composite spectra and calculating the $68\%$ confidence interval (CI; analogous to asymmetric error bars) at each wavelength.

Figure~\ref{fig:straight_stack} shows this composite spectrum (in medium blue) and the corresponding uncertainties (in light blue) over the range of rest-wavelengths with continuum S/N $\gtrsim15$, where we define the S/N as the ratio of the composite spectrum to half the $68\%$ CI. This requirement results in $\gtrsim75\%$ ($\geq17/23$~galaxies) contributing to the final composite at each wavelength. At the center of the wavelength range where the targeted auroral lines are found, the stack represents $\sim690$~object-hours of exposure time.

Aside from the ``strong" H$\alpha$, [\ion{N}{2}]$\lambda\lambda6550,6585$, and [\ion{S}{2}]$\lambda\lambda6718,6733$ lines (highlighted in the inset panel in Figure~\ref{fig:straight_stack}), no other emission lines are routinely detected in ground-based spectra of individual $z\sim2$ galaxies. Lines longward of $\sim7000$~\AA\, are virtually inaccessible from the ground at $z\gtrsim2$, due to a combination of the rising thermal background in $K$-band and decreasing atmospheric transparency. In more recent studies of high-$z$ galaxies using JWST/NIRSpec, emission lines in this wavelength range that are fainter than nebular [\ion{N}{2}] and [\ion{S}{2}] are only infrequently observed in individual galaxy spectra \citep[e.g.,][]{cameron2023,shapley2023,sanders2023}---and even these relatively strong lines are not always visible in the spectra of some distant galaxies. In the composite spectrum shown in Figure~\ref{fig:straight_stack}, we identify emission lines from eight different elements (H, He, N, O, Si, S, Ar, and Ni, denoted by the vertical dotted grey lines). Many of these have only rarely, if ever, been observed outside of the nearby universe.

\subsection{The Nebular Composite Spectrum}
\label{sec:nebular_stack}

\begin{figure*}
\centering
\includegraphics[trim={0 0.6cm 0 0},clip]{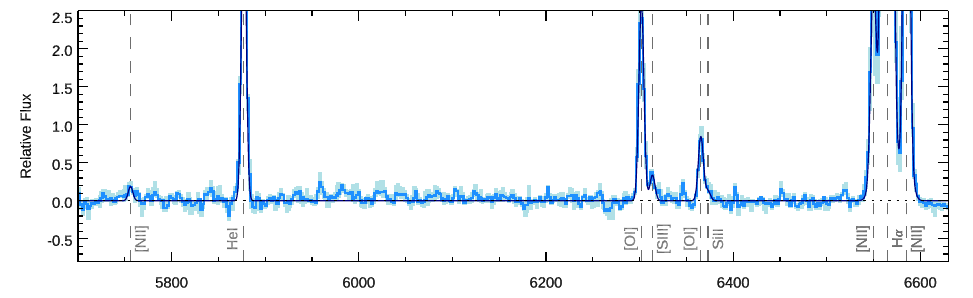}\\
\vspace{-0.07cm}
\includegraphics[trim={0 0 0 0.05cm},clip]{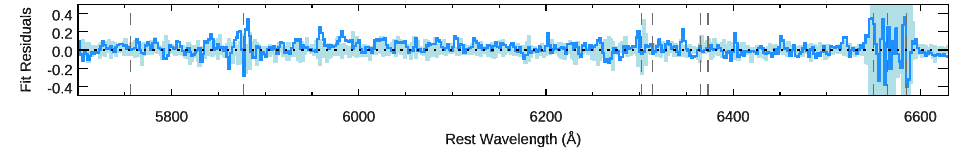}\\
\includegraphics[trim={0 0.6cm 0 0},clip]{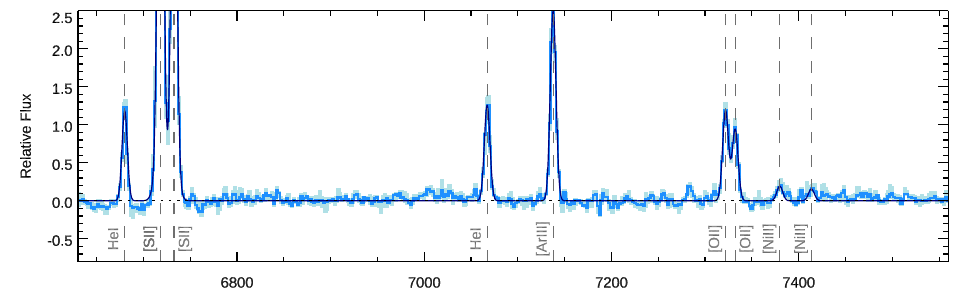}\\
\vspace{-0.07cm}
\includegraphics[trim={0 0 0 0.05cm},clip]{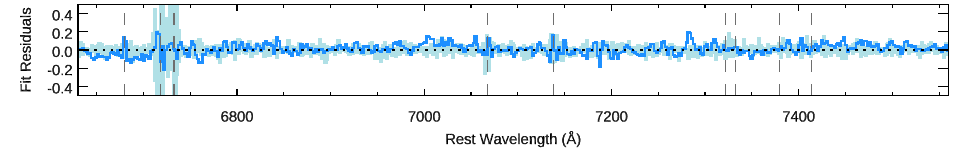}\\ 
\includegraphics[trim={0 0.6cm 0 0},clip]{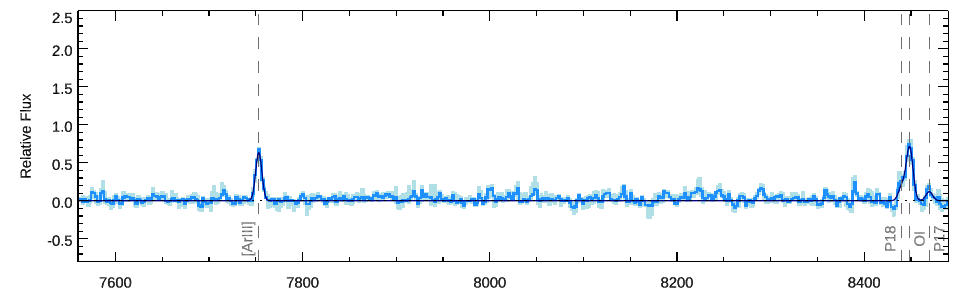}\\
\vspace{-0.07cm}
\includegraphics[trim={0 0 0 0.05cm},clip]{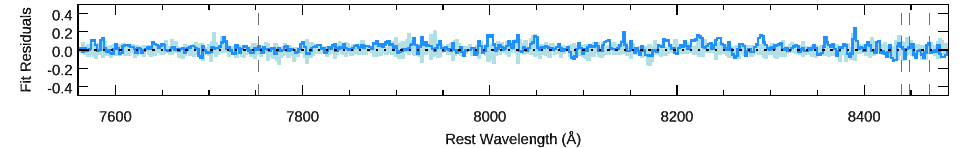} 
\caption{The nebular composite spectrum of the CECILIA sample is shown in medium blue, with the light blue shading representing the 68\% CI determined via bootstrapping. The dark blue curve shows the best-fit model, which includes emission lines of eight different elements, identified by the dashed grey lines. The strengths of these lines relative to the narrow component of H$\alpha$ are reported in Table~\ref{tab:fluxes}. The residuals from the model are shown in the bottom panels (medium blue), compared to the uncertainties on the median stack (light blue).}
\label{fig:nebular_stack}
\end{figure*}

To quantify the strength of these emission lines, we construct a second, continuum-subtracted composite spectrum. In this case, after each spectrum is flux-corrected to match the best-fit SED, the model continuum is subtracted before the spectrum is shifted into the rest frame. To remove any remaining irregular wavelength-dependent errors in the continuum subtraction, we subtract a running median, using a large window ($\Delta \lambda_{\rm rest}\sim200$~\AA) to avoid over-correcting near the emission lines. The spectra are then normalized by the measured flux in H$\alpha$ and median combined. Because H$\alpha$ falls in the detector gap for three galaxies, the nebular composite only includes 18 of the 23 galaxies used to construct the total composite spectrum. Finally, the resulting composite spectrum is converted to flux units ($\lambda F_{\lambda}$) and rescaled so that the peak flux of (narrow) H$\alpha$ is 100. Figure~\ref{fig:nebular_stack} shows this composite spectrum (in medium blue), with the flux limit chosen to facilitate inspection of the semi-strong and faint features. As before, uncertainties are estimated using bootstrapping (shown by the light blue shading). The same lines identified in Figure~\ref{fig:straight_stack} are marked by dashed grey lines here.

We determine the typical strength of these emission lines by first fitting the median composite with a model containing 73 emission lines, drawn from the catalog reported by \citet{esteban2004}, who conducted a detailed analysis of Very Large Telescope (VLT) UVES \citep{dekker2000} echelle spectrophotometry of the Orion nebula. We select those lines in the wavelength range sampled by the CECILIA nebular composite that are measured to have a flux $>0.01\%$ of H$\alpha$ in the \citet{esteban2004} spectrum. All of the lines are modeled as single Gaussians, have fixed relative wavelengths (i.e., the line centers are not allowed to move relative to one another), and are required to have the same width. For the strong [\ion{N}{2}]$\lambda\lambda6550,6585$ and semi-strong [\ion{O}{1}]$\lambda\lambda6302,6365$ doublets, which have relative strengths set by atomic physics, the ratios are fixed at 1:2.96 and 3.15:1, respectively \citep{froesefischer1983,baluja1988,tachiev2001}. A second Gaussian is included to account for broad components under the strongest lines (H$\alpha$, [\ion{N}{2}]$\lambda\lambda6550,6585$, and [\ion{S}{2}]$\lambda\lambda6718,6733$) and allowed to be offset in velocity relative to the narrow components of the same lines; all of the broad components are required to have the same line width and velocity offset. The addition of these components significantly improves the residuals from the model by accounting for excess flux detected near the H$\alpha+[$\ion{N}{2}] complex.

The 1000 bootstrap samples are fit using the same model, and the $68\%$ highest density interval (HDI) for the distributions of measured fluxes are used to determine uncertainties on the reported line fluxes. Lines are considered well detected when they have a nonzero flux in $>99\%$ of fits \emph{and} the maximum \textit{a posteriori} (MAP) value for the line flux is $>3\sigma$. Nineteen emission lines satisfy these criteria and are listed in Table~\ref{tab:fluxes}. We also include \ion{Si}{2}~$\lambda6373$ (2.9$\sigma$), the weaker [\ion{Ni}{2}] line at 7414~\AA\, (2.7$\sigma$), the Paschen line at 8470~\AA\, (3.1$\sigma$, but only nonzero in 97.5\% of the bootstrap stacks), and all of the broad components.

The dark blue curve in the top panels of Figure~\ref{fig:nebular_stack} represents the best-fit model containing the emission lines in Table~\ref{tab:fluxes}, with the fit residuals shown by the medium blue line in the bottom panels. Recall that the peak of (narrow) H$\alpha$ is set to 100 in the nebular composite, so that the peak of the semi-strong and faint emission lines corresponds to their strengths relative to the narrow component of H$\alpha$. The MAP values are also reported in Table~\ref{tab:fluxes}, along with the $68\%$ HDI for each line. Because the uncertainties are calculated via bootstrap, note that these ranges reflect contributions from both observational uncertainties on the individual line measurements and physical variation among the objects in our sample.

\section{Faint Emission Lines in High-redshift Star-forming Galaxies}
\label{sec:emlines}

\begin{deluxetable}{lcrcl}
\tablecaption{Observed Line Fluxes Relative to H$\alpha$.}
\tablenum{1}
\tablehead{\colhead{Ion} & \colhead{$\lambda_{\rm vac}$} & \colhead{F$(\lambda)$} & \colhead{Range} & \colhead{Notes} \\
\colhead{} & \colhead{(\AA)} & \colhead{(\%)} & \colhead{(\%)} & \colhead{}} 
\startdata
\multicolumn{5}{c}{Narrow components} \\
\hline
${\rm [N~II]     }$ & 5756.24 & $0.20$ & $0.15-0.25$ & Auroral line \\
${\rm He~I       }$ & 5877.27 & $4.12$ & $3.74-4.48$ & \\
${\rm [O~I]      }$ & 6302.04 & $2.80$ & $2.36-3.08$ & \\
${\rm [S~III]    }$ & 6313.85 & $0.34$ & $0.29-0.41$ & Auroral line \\
${\rm [O~I]      }$ & 6365.54 & $0.91$ & $0.77-1.01$ & \\
${\rm Si~II      }$ & 6373.12 & $0.12$ & $0.09-0.21$ & \\
${\rm [N~II]     }$ & 6549.86 & $2.55$ & $1.86-3.02$ & \\
${\rm H\alpha   }$ & 6564.62 & $100.00$ & \nodata\nodata & \\
${\rm [N~II]     }$ & 6585.27 & $7.59$ & $5.54-8.99$ & \\
${\rm He~I       }$ & 6679.99 & $1.21$ & $1.09-1.34$ & \\
${\rm [S~II]     }$ & 6718.29 & $7.33$ & $5.55-7.76$ & \\
${\rm [S~II]     }$ & 6732.68 & $6.49$ & $4.94-6.93$ & \\
${\rm He~I       }$ & 7067.23 & $1.25$ & $1.07-1.51$ & \\
${\rm [Ar~III]   }$ & 7137.75 & $2.63$ & $2.22-2.86$ & \\
${\rm [O~II]     }$ & 7321.94 & $1.21$ & $1.10-1.38$ & Auroral line \\
${\rm [O~II]     }$ & 7332.21 & $0.95$ & $0.88-1.12$ & Auroral line \\
${\rm [Ni~II]    }$ & 7379.86 & $0.22$ & $0.15-0.30$ & \\
${\rm [Ni~II]    }$ & 7413.65 & $0.17$ & $0.11-0.24$ & \\
${\rm [Ar~III]   }$ & 7753.23 & $0.64$ & $0.57-0.72$ & \\
${\rm P18      }$ & 8440.28 & $0.20$ & $0.13-0.27$ & \\
${\rm O~I        }$ & 8448.57 & $0.73$ & $0.64-0.87$ & \\
${\rm P17      }$ & 8469.58 & $0.13$ & $0.09-0.18$ & \\
\hline
\multicolumn{5}{c}{Broad components} \\
\hline
${\rm [N~II]     }$ & 6549.86 & $0.86$ & $0.15-1.56$ & \\
${\rm H\alpha   }$ & 6564.62 & $12.63$ & $\phn6.01-28.31$ & \\
${\rm [N~II]     }$ & 6585.27 & $2.55$ & $0.45-4.65$ & \\
${\rm [S~II]     }$ & 6718.29 & $0.95$ & $0.00-2.48$ & \\
${\rm [S~II]     }$ & 6732.68 & $0.81$ & $0.00-1.50$ &
\enddata
\tablecomments{[F(H$\alpha_{\rm narrow})=100$]}
\label{tab:fluxes}
\end{deluxetable}

In this section, we highlight individual semi-strong ($\approx2-3$\% of H$\alpha$) and faint ($\lesssim1$\% of H$\alpha$) emission lines detected in the CECILIA composite spectra and briefly comment on how they may be used to study high-$z$ galaxies.

\subsection{Auroral Lines}

Of all the faint lines present in the rest-optical spectra of star-forming regions, the auroral\footnote{``Auroral" lines are forbidden transitions from the second excited state to the first excited state of ions of heavy elements and can be paired with observations of the corresponding ``nebular" (first excited state to ground state) lines to determine $T_e$ in low-density gas where collisional de-excitation does not play a significant role.} lines that can be used to implement the direct method of measuring metallicities have received the most attention in studies of high-$z$ galaxies. Foremost among these is [\ion{O}{3}]$\lambda4364$, which falls at $\lambda_{\rm obs} \approx 1.3-1.7~\mu$m for the $z\sim2-3$ CECILIA galaxies. As described in Section~\ref{sec:predictions}, CECILIA instead targets two auroral lines at longer wavelengths that are not only predicted to be stronger than [\ion{O}{3}]$\lambda4364$, but also fall at $\lambda_{\rm obs}\gtrsim2.0~\mu$m, where JWST is more sensitive. In total, three auroral lines fall in the wavelength range sampled by the composite spectra shown in Figures~\ref{fig:straight_stack} and \ref{fig:nebular_stack} and form the basis of our discussion here: [\ion{N}{2}]$\lambda5756$, [\ion{S}{3}]$\lambda6313$ [\ion{O}{2}]$\lambda\lambda7322,7332$.

\begin{figure*}
\centering
\includegraphics[height=5.5cm]{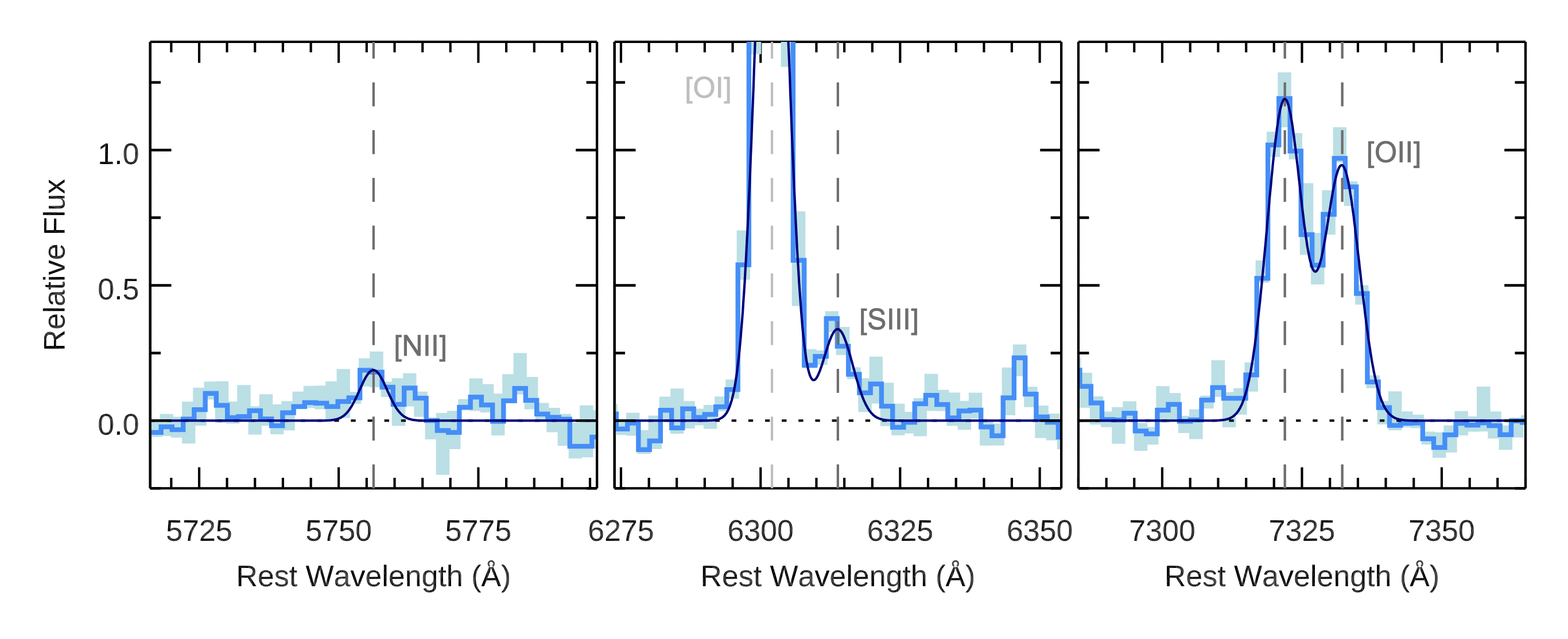}
\vspace{-0.15in}
\caption{The three auroral lines observed in the CECILIA stack are shown in separate panels with the same flux scale, in the same manner as Figure~\ref{fig:nebular_stack}. Auroral [\ion{N}{2}]$\lambda5756$ (left panel) is the weakest of the three and is estimated to be $0.15-0.25\%$ the strength of H$\alpha$. Both [\ion{S}{3}]$\lambda6313$ (center panel) and [\ion{O}{2}]$\lambda7322,7332$ (right panel) are noticeably stronger, and the oxygen lines are clearly the brightest auroral features in the $\lambda=5700-8500$~\AA\, range, with each being $\sim1\%$ the strength of H$\alpha$.}
\label{fig:auroral_lines}.
\end{figure*}

The strongest of these is [\ion{O}{2}]$\lambda\lambda7322,7332$, with both lines observed to be $\sim1\%$ the strength of the narrow component of H$\alpha$ (right panel of Figure~\ref{fig:auroral_lines}). \citet{sanders2023o2} recently reported the detection of this feature (actually a quadruplet) in two $z=2.18$ galaxies, which had each been observed for $\sim15$~hr using Keck/MOSFIRE. Using their measurements to calculate direct-method oxygen abundances, they find moderate $12+\log(\rm{O/H})=7.89\pm0.20$ and $12+\log(\rm{O/H})=8.24\pm0.27$. Comparing these abundances to the predictions in the bottom panel of Figure~\ref{fig:auroral}, we see that they lie near the broad peak of the predicted line strengths and, thus, likely represent only the ``tip of the iceberg": other deep spectroscopic studies should uncover auroral [\ion{O}{2}] lines in galaxies with a wider range of O/H. This is one of the main goals of the CECILIA program (Section~\ref{sec:predictions}), which includes many high-confidence detections of these lines in individual galaxy spectra that will be investigated in a subsequent paper. 

The other auroral line specifically targeted by CECILIA is [\ion{S}{3}]$\lambda6313$ (middle panel of Figure~\ref{fig:auroral_lines}), which samples a higher ionization zone than auroral [\ion{O}{2}]. Whereas this line is routinely used to measure abundances in nearby extragalactic \ion{H}{2} regions, it has never been reported in observations of galaxies outside the local universe. It is significantly detected in the CECILIA composites, and we have preliminary evidence of its presence in spectra of individual CECILIA galaxies. However, because of its faintness ($0.29-0.41\%$ the strength of H$\alpha$) and proximity to the comparatively stronger [\ion{O}{1}]$\lambda6302$ line, [\ion{S}{3}]$\lambda6313$ is unlikely to be accessible in shallower or lower resolution JWST observations of objects similar to the CECILIA sample.

\begin{figure*}
\centering
\includegraphics[height=5.5cm]{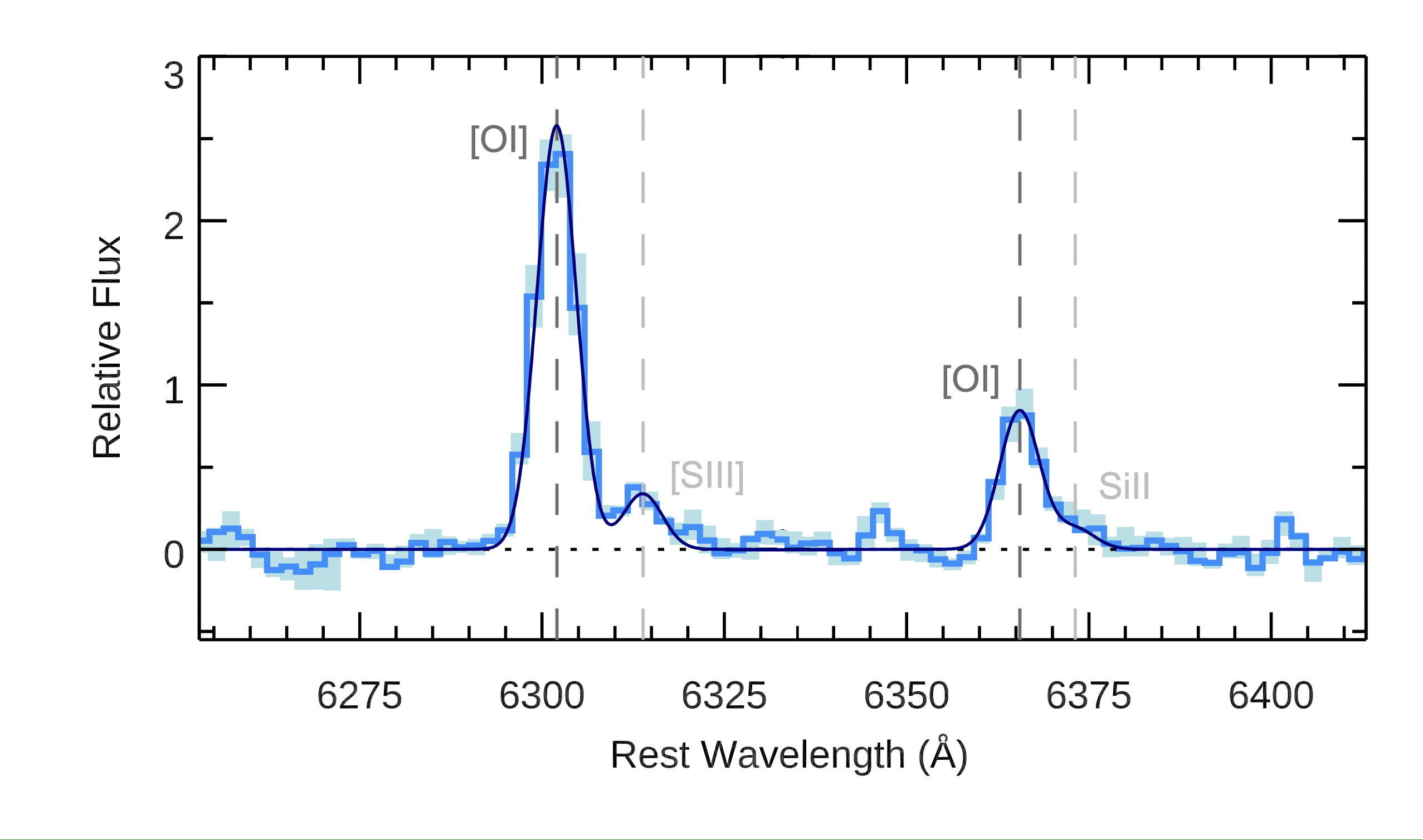}
\includegraphics[height=5.5cm]{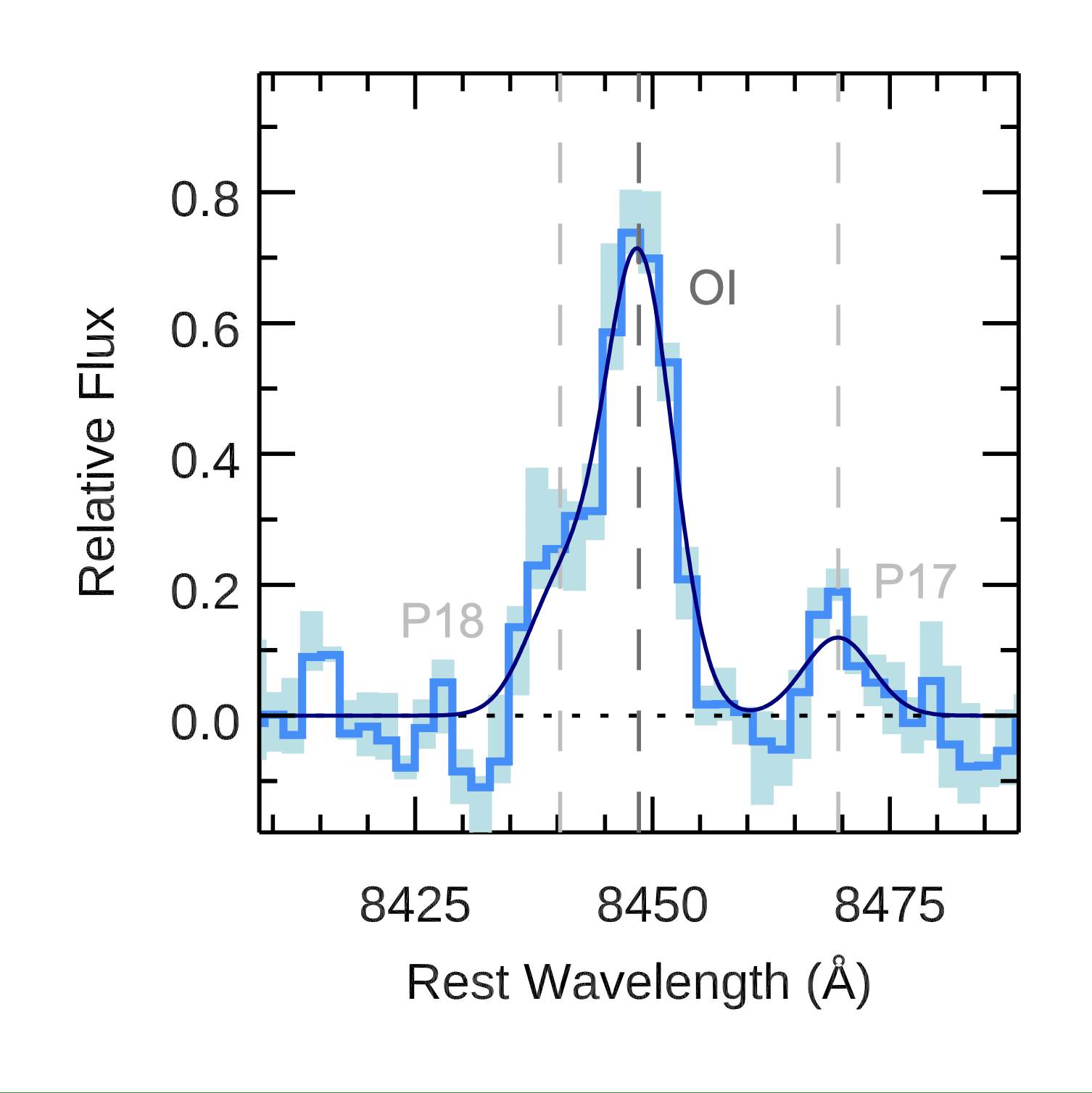}\\
\vspace{-0.15in}
\includegraphics[height=5.5cm]{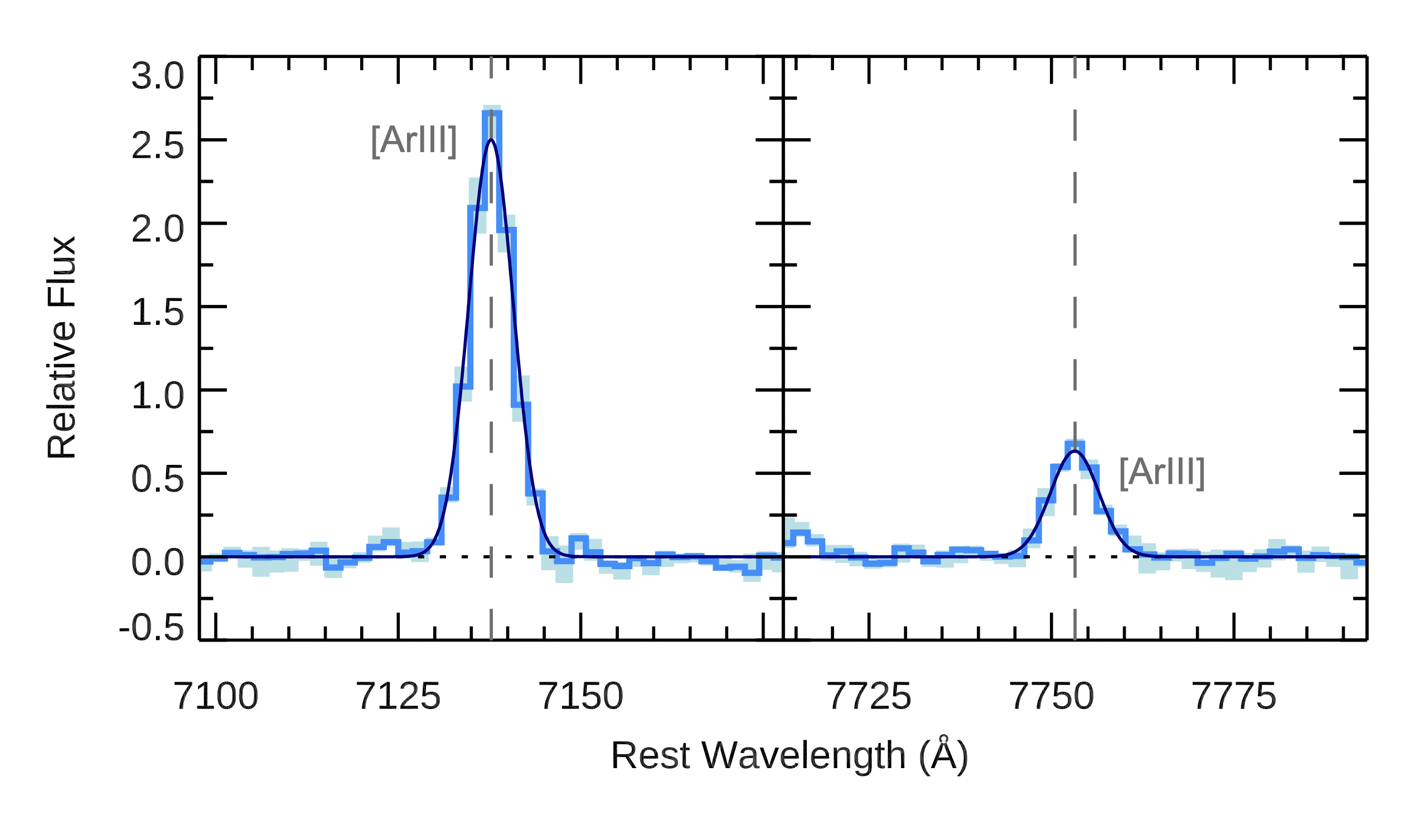}
\includegraphics[height=5.5cm]{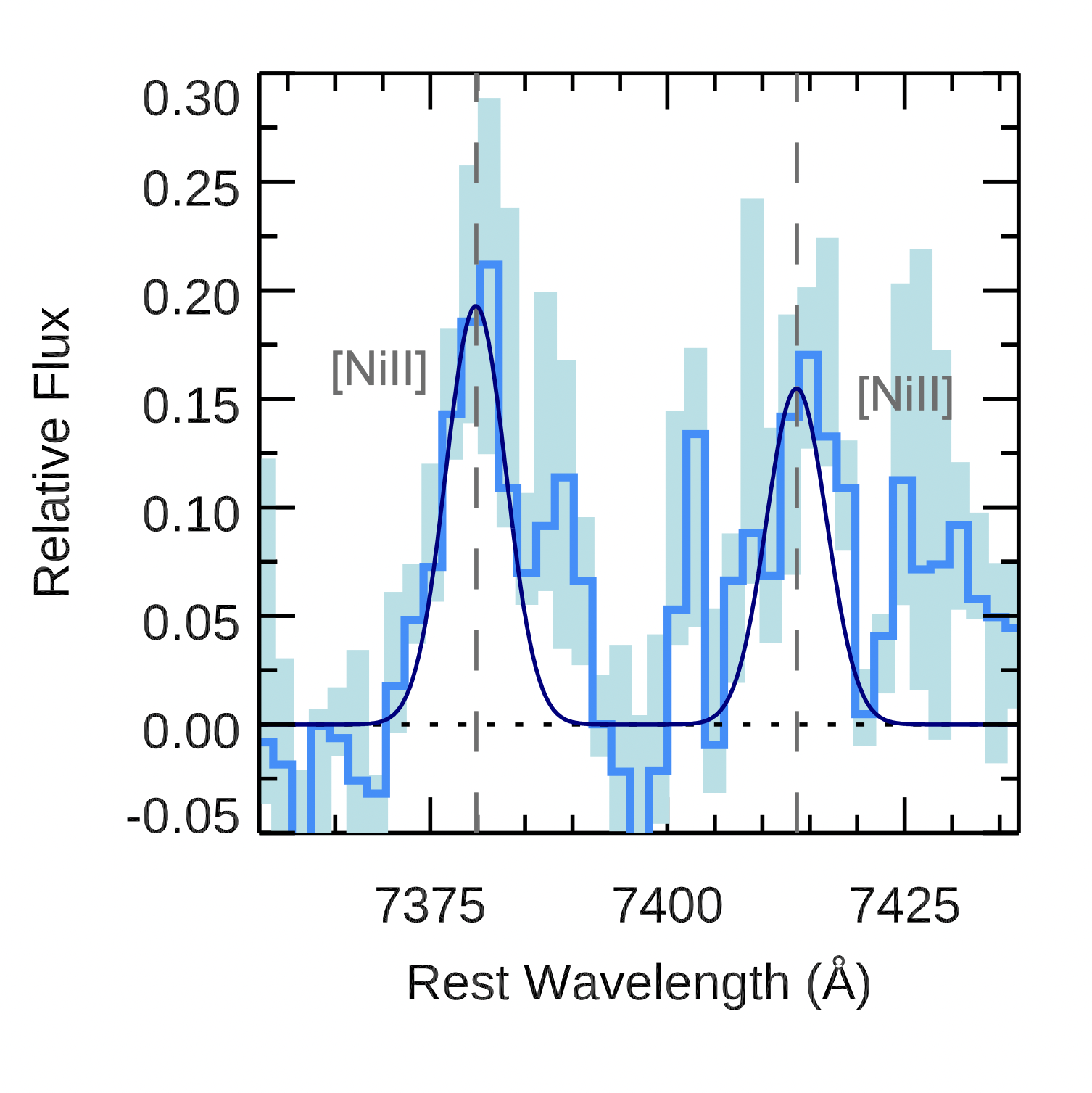}
\vspace{-0.15in}
\caption{Semi-strong and faint emission lines of four elements---O, Si, Ar, and Ni---are shown in the same manner as Figures~\ref{fig:nebular_stack} and \ref{fig:auroral_lines}, but with a different flux range used each panel. The top row shows forbidden [\ion{O}{1}]$\lambda6302,6565$ (upper left), with [\ion{S}{3}]$\lambda6313$ and \ion{Si}{2}~$\lambda6373$ in the red wings of each line, respectively, and permitted \ion{O}{1} at $8449$~\AA, blended with Pa18 (upper right). The bottom row shows forbidden lines of [\ion{Ar}{3}]$\lambda\lambda7138,7753$ (lower left) and [\ion{Ni}{2}]$\lambda7380,7414$ (lower right). The stronger [\ion{O}{1}] and [\ion{Ar}{3}] lines have now been observed in deep spectra of individual high-$z$ galaxies \citep[e.g.,][]{cameron2023,sanders2023o2}, but the permitted \ion{O}{1}~$\lambda8449$ line and forbidden [\ion{Ni}{2}] are only rarely observed, even in observations of nearby galaxies and \ion{H}{2} regions.}
\label{fig:otherlines}
\end{figure*}

The third auroral line observed in the composite spectra is [\ion{N}{2}]$\lambda5756$, at $0.15-0.25\%$ the strength of H$\alpha$ (left panel of Figure~\ref{fig:auroral_lines}). It is formally detected at 4.6$\sigma$ in the stack but is likely too faint to be detected in the individual spectra of high-$z$ galaxies, even using long exposure times with JWST. Still, for the sample of galaxies where it \emph{is} possible to detect, it may serve as an important tool for calibrating the $T_e$ relation between different ionization zones \citep[e.g.,][]{garnett1992,esteban2009,croxall2016,yates2020,rogers2021}.

We use our measurement of [\ion{N}{2}]$\lambda5756$ to calculate $T_e$ and the corresponding direct-method ionic abundance. First, we use the line strengths for [\ion{S}{2}]$\lambda\lambda6718,6733$ to determine the electron density and find $n_e\approx285$~cm$^{-3}$, which is consistent with values previously reported for KBSS galaxies \citep{strom2017} and other $z\sim2-3$ galaxy samples \citep[e.g.,][]{sanders2016}. This density is then combined with the measurements of nebular and auroral [\ion{N}{2}] to calculate $T_e$ using the \texttt{PyNeb} package \citep{luridiana2015}. However, because the nebular stack does not contain both H$\alpha$ and H$\beta$, which are required to determine the Balmer decrement and robustly constrain the reddening, we adopt the interquartile range in E(B$-$V) for the KBSS parent sample, E(B$-$V)$=0.06-0.47$ \citep{strom2017}. Using these values, we find $T_e$[\ion{N}{2}]$=13630\pm2540$~K, where the reported uncertainties also capture the likely range in reddening for $z\sim2-3$ galaxies. This temperature can be used to infer the abundances of low-ionization species, and we ultimately calculate $12+\log($N$^+$/H$^+)=6.33^{+0.18}_{-0.30}$ and $12+\log($S$^+$/H$^+)=5.70^{+0.16}_{-0.26}$ using the nebular lines for both ions; because the auroral [\ion{O}{2}]$\lambda\lambda7322,7332$ lines are, by definition, strong functions of $T_e$, any O$^+$/H$^+$ abundance determined using $T_e$[\ion{N}{2}] would have much larger uncertainties. We cannot confidently determine the contribution from other common (but unseen) ionization states of N and S using the nebular composite alone and so do not report total abundances here, but we plan to revisit the issue of appropriate ionization correction factors for high-$z$ galaxies in future work.

\subsection{Other Semi-strong and Faint Lines}

In addition to the three auroral emission lines, we also detect eight semi-strong and faint emission lines from four different heavy elements. Cutouts of the nebular composite spectrum near these features are shown in Figure~\ref{fig:otherlines}, in the same manner as Figures~\ref{fig:nebular_stack} and \ref{fig:auroral_lines}.

The strongest of the lines is [\ion{O}{1}]$\lambda6302$ (upper left panel of Figure~\ref{fig:otherlines}), which is observed to be $2.36-3.08\%$ the strength of the narrow component of H$\alpha$ and is significantly stronger than the auroral [\ion{S}{3}]$\lambda6313$ line in its red wing. Its partner line at 6365~\AA\, is $\approx3.15\times$ weaker as set by atomic physics and is blended with \ion{Si}{2}~$\lambda6373$, which also probes mostly neutral and low-ionization gas. Rather than being an abundance diagnostic, [\ion{O}{1}]$\lambda6302$ is most commonly used as a way to identify the principal ionization mechanism in emission line galaxies using a form of the Baldwin-Philips-Terlevich (BPT) diagram \citep{baldwin1981,veilleux1987}. It can also be useful for identifying contributions from diffuse ionized gas and shocks \citep[e.g.,][]{tullmann2000,moy2002,zhang2017}. Although widely studied in the local universe, including in large samples such as the Sloan Digital Sky Survey \citep[SDSS;][]{kewley2006,law2021}, [\ion{O}{1}]$\lambda6302$ is not commonly reported for individual $z\gtrsim2$ galaxies. More recently, however, it has been observed in a handful of high-$z$ galaxies with moderately deep ground-based or JWST spectroscopy \citep[e.g.,][]{cameron2023,clarke2023,sanders2023o2}. Given its typical strength, this semi-strong line should provide an accessible and promising method for discriminating between AGN and star formation and probing low-ionization gas in high-$z$ galaxies.

The lower left panel of Figure~\ref{fig:otherlines} shows the widely-spaced [\ion{Ar}{3}]$\lambda\lambda7138,7753$ lines, which, like lines of O$^{++}$, trace the gas in the high ionization zone of star-forming regions. In low-$z$ galaxies, the stronger [\ion{Ar}{3}]$\lambda7138$ line has been used to determine absolute argon abundances and relative abundance ratios, such as Ar/O \citep[e.g.,][]{berg2015,croxall2016,rogers2021}, after accounting for unseen ionization states of Ar. At $2.22-2.86$\% the strength of H$\alpha$, comparable to [\ion{O}{1}]$\lambda6302$, this line is one of the strongest heavy metal lines present in the CECILIA nebular composite spectrum, aside from the familiar strong lines; \citet{sanders2023o2} find similar ratios for their two $z=2.18$ galaxies. Thus, [\ion{Ar}{3}]$\lambda7138$ is also an attractive target for spectroscopic studies of galaxy enrichment. Although both Ar and O are nominally produced by the same mechanism in massive stars, differences in Ar/O as a function of overall enrichment could reflect a dependence of stellar nucleosynthesis on metallicity \citep[e.g.,][]{kennicutt2003,izotov2006}.

Permitted \ion{O}{1} at 8449~\AA\, (upper right panel of Figure~\ref{fig:otherlines}) is one of the most commonly used recombination lines for measuring metallicity in astrophysical environments where such transitions can be observed \citep{maiolino2019}. Typically, however, metal recombination lines are too weak to be useful diagnostics, even in $z\sim0$ galaxies and \ion{H}{2} regions, so its presence in the CECILIA composite is unexpected. We consulted the database maintained by the Atomic Spectroscopy Data Center at the National Institute of Standards and Technology (NIST), but no other likely candidates for emission lines at the same rest wavelength were identified. \ion{O}{1}~$\lambda8449$ is blended with the Paschen series line at 8440~\AA\, (P18) in its left wing, and the neighboring Paschen line at 8470~\AA\, (P17) is also detected in the nebular composite, relieving concerns that poor wavelength calibration may have led to misidentifying the line. The anomalous strength of this line in the Orion Nebula is in fact a longstanding puzzle \citep[e.g.,][]{morgan1971,danziger1974}, but \citet{grandi1975} showed that direct excitation by starlight is the most likely origin, rather than recombination from O$^+$ or Ly$\beta$ fluorescence. The strength of \ion{O}{1} $\lambda8449$ in the CECILIA composite relative to H$\alpha$ ($0.64-0.87$\%) is comparable to that reported for the Orion Nebular \citep[$0.56$\%;][]{esteban2004}, suggesting that the same mechanism may also be dominant in high-$z$ galaxies. If true, the contribution from direct excitation would severely limit the utility of \ion{O}{1} $\lambda8449$ as a metallicity tracer; \citet{garcia-rojas2006} and \citet{garcia-rojas2007} used both \ion{O}{1}~$\lambda7771$ and \ion{O}{1} $\lambda8449$ to derive O$^+$ abundances in Galactic \ion{H}{2} regions but found an order of magnitude larger O$^+$/H$^+$ using \ion{O}{1} $\lambda8449$, likely due to this effect.

Also puzzling is the detection of [\ion{Ni}{2}]$\lambda\lambda7380,7414$ (shown in the lower right panel of Figure~\ref{fig:otherlines}), which are $0.15-0.30$\% and $0.11-0.24$\% the strength of H$\alpha$, respectively. There are comparatively few references to the observation of this line in astrophysical objects, but a handful of studies have reported measurements of [\ion{Ni}{2}]$\lambda7380$ and the corresponding Ni$^+$ abundances in gaseous nebulae in the Milky Way \citep{dennefeld1982,fesen1982,henry1988,esteban1999}. In many of these cases, [\ion{Ni}{2}]$\lambda7380$ was seen to be much stronger than expected relative to the associated [\ion{Ni}{2}]$\lambda7414$ line, which is only marginally detected in the CECILIA composite spectra. Other authors have explained this by invoking fluorescence by the UV continuum, similar to [\ion{Fe}{2}] lines \citep{lucy1995}, and/or collisional excitation in very high density ($n_e\approx10^6$~cm$^{-3}$) gas \citep{bautista1996}. Similar to \ion{O}{1}~$\lambda8449$, no reasonable alternatives were identified in the NIST database, but without significantly detecting [\ion{Fe}{2}] lines that may also be impacted by the same physical mechanism, it is difficult to speculate about its appearance here.

\subsection{\texorpdfstring{Broad H$\alpha$}{Broad H-alpha}}

\begin{figure}
\centering
\includegraphics[width=8.8cm]{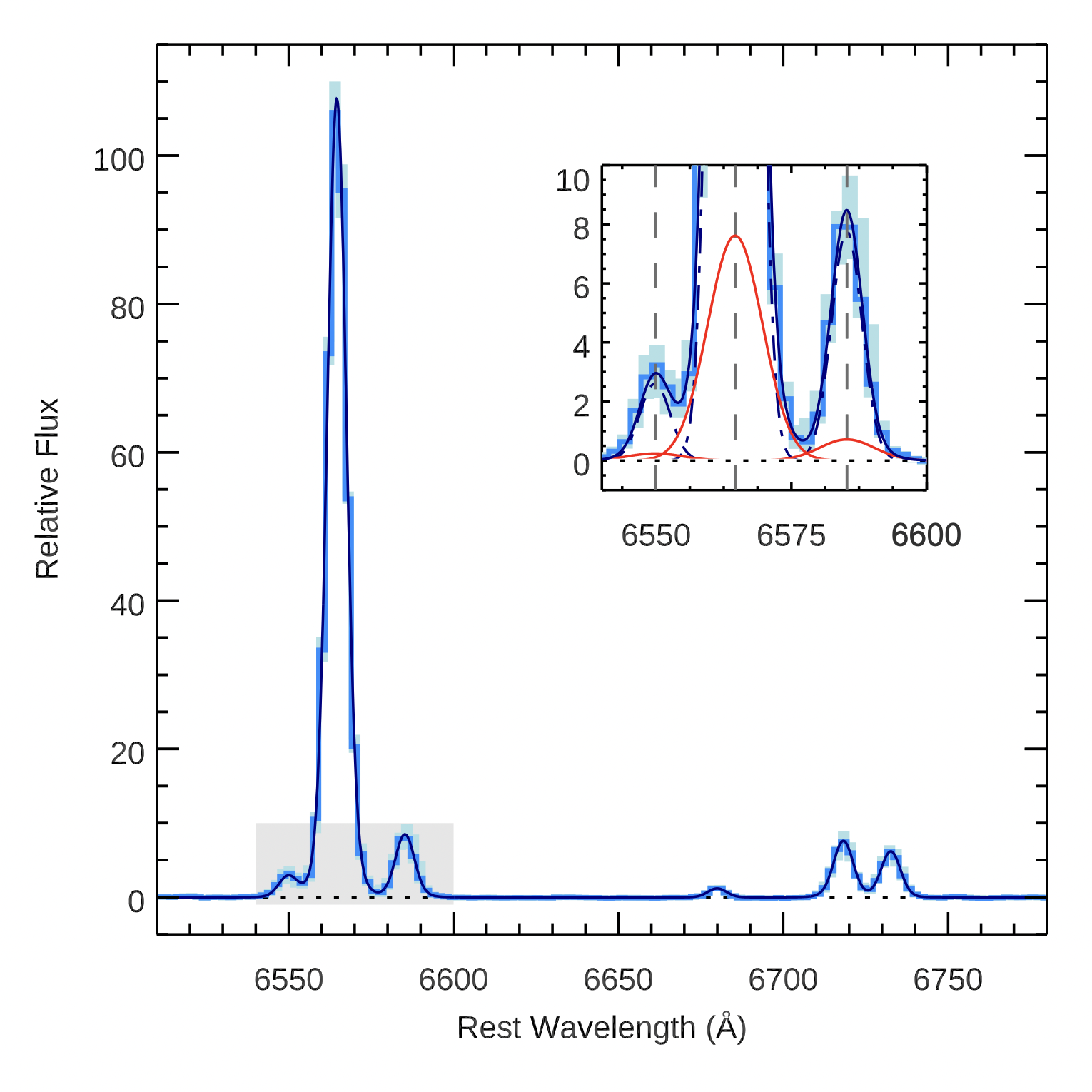}
\caption{The same nebular composite spectrum from Figure~\ref{fig:nebular_stack} is shown near the strong H$\alpha$, [\ion{N}{2}]$\lambda\lambda6550,6585$, and [\ion{S}{2}]$\lambda\lambda6718,6733$ lines, with a more extended flux range to show the peaks of the lines. The inset panel is a zoom-in on the grey shaded region, highlighting the broad components of H$\alpha$ and [\ion{N}{2}] (in red). The narrow components of all three lines are illustrated by the dot-dashed navy curves.}
\label{fig:ha_zoom}
\end{figure}

Broad line emission has been observed in both spectra of individual galaxies and composite spectra of galaxies at $z\sim2$ and is usually attributed to galaxy-scale ionized gas outflows \citep[see Section 4.6 of the review by][and references therein]{forster-schreiber2020}; in contrast, the frequently brighter, narrow components of emission lines trace galaxies' star-forming regions. Both active galactic nuclei \citep[AGN; e.g.,][]{nesvadba2008,genzel2014,forster-schreiber2014,cresci2015} and star-formation \citep[e.g.,][]{genzel2011,davies2019,freeman2019} can generate these outflows, resulting in differences in inferred outflow velocity (i.e., emission line width), with AGN typically driving higher velocity outflows than feedback from massive stars.

In order to achieve a good fit to the nebular composite spectrum, we include two Gaussian components for the strongest lines to account for excess flux that results in large residuals from a model with only a single (narrow) component. Based on the results from fitting the 1000 bootstrap samples, these broad components have a FWHM of $536^{+45}_{-167}$~km~s$^{-1}$ and are consistent with no velocity offset relative to the narrow components, which have a FWHM of $288^{+15}_{-20}$~km~s$^{-1}$. The broad H$\alpha$ line is $6.01-28.31$\% the strength of the narrow component of H$\alpha$, with significantly weaker broad components observed for nebular [\ion{N}{2}] and [\ion{S}{2}]; this is consistent with the low end of the range reported for a similar sample of $z\sim2$ star-forming galaxies \citep{freeman2019}. If these components in the CECILIA composite do indeed reflect the presence of ionized gas outflows, the evidence for broad (albeit weak) line emission in forbidden transitions and the moderate velocity width suggest that they are likely driven by star formation. Comparable FWHM velocities of $\sim400-500$~km~s$^{-1}$ are observed in deep VLT/SINFONI \citep{eisenhauer2003,bonnet2004} spectra of star-forming clumps at $z\sim2$ \citep[e.g.,][]{newman2012clump,forster-schreiber2019}. However, because of the additional median filtering required to remove remaining fluctuations in the continuum of individual galaxy spectra (Section~\ref{sec:nebular_stack}), the detailed properties of any broad line emission in the CECILIA stack could be systematically biased.

\section{Conclusions}
\label{sec:conclusions}

We have reported the first results from the CECILIA program (JWST PID 2593), which obtained ultra-deep $\sim30$~hr NIRSpec/G235M observations of 33 star-forming galaxies at $z\sim1-3$. Using data for 23 of these galaxies, we constructed rest-optical composite spectra, both with and without the stellar continuum, corresponding to exposure times of 690~object-hours and 540~object-hours, respectively. These composites, shown in Figures~\ref{fig:straight_stack} and \ref{fig:nebular_stack}, provide one of the most detailed views to date of star-forming galaxies in the early universe and function as an atlas of their characteristic rest-optical emission line spectra.

The principal findings based on our analysis of the stacked spectra are as follows:
\begin{itemize}
\item We significantly detect emission lines of eight different elements (H, He, N, O, Si, S, Ar, and Ni), including evidence for broad line emission under H$\alpha$, [\ion{N}{2}]$\lambda\lambda6550,6585$, and [\ion{S}{2}]$\lambda\lambda6718,6733$. The strengths of these lines relative to the narrow component of H$\alpha$ are reported in Table~\ref{tab:fluxes}. 
\item Aside from strong [\ion{N}{2}], H$\alpha$, and [\ion{S}{2}], which have previously been studied in large ground-based spectroscopic samples, the majority of emission lines are $\lesssim3\%$ the strength of H$\alpha$. Some of these features, such as [\ion{O}{1}]$\lambda6302$ (shown in the upper left panel of Figure~\ref{fig:otherlines}), are now being detected in JWST spectra of individual high-$z$ galaxies, and we expect other lines with strengths $\gtrsim2-3$\% that of H$\alpha$ to be good candidates for spectroscopic follow-up of large samples. In addition to the stronger forbidden [\ion{O}{1}] line, these semi-strong lines include the \ion{He}{1} line at $\lambda5877$ and [\ion{Ar}{3}]$\lambda7138$ (shown in the bottom left panel of Figure~\ref{fig:otherlines}).
\item The three auroral emission lines present at $\lambda_{\rm rest}\approx5700-8500$ ([\ion{N}{2}]$\lambda5756$, [\ion{S}{3}]$\lambda6313$, [\ion{O}{2}]$\lambda\lambda7322,7332$, shown in Figure~\ref{fig:auroral_lines}) are $\lesssim1\%$ the strength of H$\alpha$. Using our measurements of auroral and nebular [\ion{N}{2}], we find $T_e$[\ion{N}{2}]~$=13630\pm2540$~K, which is the first time a $T_e$ has been reported for high-redshift galaxies using this tracer. Although we have not reported the significance of detections in individual galaxy spectra in this work, it seems likely that these auroral lines will remain out of reach of typical observations of high-$z$ galaxies, particularly those with low SFRs, low ionization, and/or high metallicity. This only underscores the need for more accurate line ratio diagnostics for metallicity that make use of the strong and semi-strong emission lines present in galaxies' rest-optical spectra.
\item We measure broad ($536^{+45}_{-167}$~km~s$^{-1}$ FWHM) line emission under the strongest lines and a broad component of H$\alpha$ that is $6.01-28.31$\% the strength of the narrow component (Figure~\ref{fig:ha_zoom}). These results appear indicative of star-formation driven outflows. However, we caution that, owing to remaining uncertainties in the flux calibration (see the discussion in Section~\ref{sec:reduction}), the appearance of this component should not be over-interpreted. We defer a more detailed discussion of broad line emission and its connection to galaxy outflows to a future paper.
\end{itemize}

JWST is delivering on its promise to provide access to faint emission lines in the spectra of $\gtrsim2$ galaxies, evidenced not only by what we have presented in this let, but also by the many exciting results based on NIRSpec/MSA and NIRCam grism spectroscopy that have been published over the last year. Deep observations, such as those obtained as part of CECILIA and outlined here, will be critical for developing and testing the new tools necessary to accurately interpret this wealth of data. As known issues with JWST data products continue to be resolved, it will benefit the extragalactic community to revisit some of the earliest observations---with the benefit of hindsight and these new tools---in order to maximize the scientific impact of these data. To aid in this effort, forthcoming work with CECILIA will focus on (1) $T_e$ measurements and direct-method metallicities for the sample of galaxies introduced here, as well as (2) new line ratio diagnostics for gas-phase oxygen abundance.

\begin{acknowledgements}
The authors thank Jane Rigby, Taylor Hutchison, and Marcia Rieke for their advice regarding the reduction of the JWST data, as well as Jenny Greene for her input on the scope of the discussion. We are also grateful to the JWST/NIRSpec team for their ongoing work to support this complex and powerful instrument.

ALS, GCR and RFT acknowledge partial support from the JWST-GO-02593.008-A, JWST-GO-02593.004-A, and JWST-GO-02593.006-A grants, respectively. RFT also acknowledges support from the Pittsburgh Foundation (grant ID UN2021-121482) and the Research Corporation for Scientific Advancement (Cottrell Scholar Award, grant ID 28289).

This work is primarily based on observations made with NASA/ESA/CSA JWST, associated with PID 2593, which can be accessed via doi:\dataset[10.17909/x66z-p144]{https://doi.org/10.17909/x66z-p144}. The data were obtained from the Mikulski Archive for Space Telescopes (MAST) at the Space Telescope Science Institute, which is operated by the Association of Universities for Research in Astronomy, Inc., under NASA contract NAS 5-03127 for JWST.

Some of the data used to generate the original line flux predictions were obtained at the W.M. Keck Observatory, which is operated as a scientific partnership between the California Institute of Technology, the University of California, and NASA. The Observatory was made possible by the generous financial support of the W.M. Keck Foundation, and the authors wish to recognize and acknowledge the significant cultural role and reverence that the summit of Maunakea has within the indigenous Hawaiian community.
\end{acknowledgements}

\facilities{JWST (NIRSpec)}
\software{BPASSv2 \citep{stanway2016,eldridge2017}, Cloudy \citep{ferland2013}, \texttt{GalDNA} \citep{strom2018}, JWST Calibration Pipeline \citep{calwebb_v1.10.0}, \texttt{grizli} \citep{grizli}, \texttt{msaexp} \citep{msaexp}, \texttt{PyNeb} \citep{luridiana2015}}

\bibliography{strom_ref_library}
\bibliographystyle{aasjournal}

\end{CJK*}
\end{document}